%% file: 00_main.tex
\documentclass[sigchi]{acmart}
\AtBeginDocument{%
  }

\setcopyright{acmlicensed}
\copyrightyear{2018}
\acmYear{2018}
\acmDOI{XXXXXXX.XXXXXXX}
\acmConference[Conference acronym 'XX]{Make sure to enter the correct
  conference title from your rights confirmation email}{June 03--05,
  2018}{Woodstock, NY}
\acmISBN{978-1-4503-XXXX-X/2018/06}

\newif\ifsubmit
\submitfalse
\ifsubmit
    \newcommand{\vincent}[1]{}
    \newcommand{\hari}[1]{}
\else
\newcommand{\vincent}[1]{\textcolor[rgb]{.1, .8, .1}{~VC: #1}}
\newcommand{\hari}[1]{{\textcolor{purple}{\bf [*** HS: #1]}}}
\fi

\usepackage{xspace}
\usepackage{multirow}
\usepackage{xcolor}
\usepackage{tcolorbox}
\tcbuselibrary{most}
\usepackage{float}
\usepackage{colortbl}
\usepackage{tikz}
\usepackage{xcolor}
\usetikzlibrary{arrows.meta, positioning, fit, backgrounds, calc, shapes.geometric, matrix}

\newtcbox{\macrotag}[1][]{on line, arc=3pt, outer arc=3pt, colback=blue!12, colframe=blue!12, boxsep=0pt, left=3pt, right=3pt, top=1.5pt, bottom=1.5pt, fontupper=\scriptsize\sffamily, #1}
\newtcbox{\microtag}[1][]{on line, arc=3pt, outer arc=3pt, colback=orange!18, colframe=orange!18, boxsep=0pt, left=3pt, right=3pt, top=1.5pt, bottom=1.5pt, fontupper=\scriptsize\sffamily, #1}

\newcommand{\system}{Tellimation\xspace} 
\newcommand{\systems}{Tellimation's\xspace}

\begin{document}
\title{\system: Making Narrative Gaps Visible in Children’s Storytelling with Just-in-Time Animation}

\author{Vincent Cavez}
 \affiliation{%
  \institution{Stanford University}
   \country{USA}
 }
 \email{vcavez@stanford.edu}

\author{Marielle Zheng}
 \affiliation{%
  \institution{Stanford University}
   \country{USA}
 }
 \email{mxmzheng@stanford.edu}

\author{Momin Siddiqui}
 \affiliation{%
  \institution{Georgia Institute of Technology}
   \country{USA}
 }
 \email{momin@gatech.edu}

\author{Abbie L Olszewski}
 \affiliation{%
  \institution{University of Nevada, Reno}
   \country{USA}
 }
\email{aolszewski@unr.edu}

\author{Srirangaraj Setlur}
 \affiliation{%
  \institution{University of Buffalo}
   \country{USA}
 }
 \email{setlur@buffalo.edu}

\author{Maneesh Agrawala}
 \affiliation{%
  \institution{Stanford University}
   \country{USA}
 }
 \email{maneesh@cs.stanford.edu}

\author{Hari Subramonyam}
 \affiliation{%
  \institution{Stanford University}
   \country{USA}
 }
 \email{harihars@stanford.edu}

\renewcommand{\shortauthors}{Cavez, et al.}

\begin{abstract}

Illustrated scenes are used to prompt children's stories, yet children with language difficulties often omit the details, actions, and relationships that make a story coherent. We introduce \system, which animates scene elements a child has omitted or misdescribed, drawing attention to a narrative gap without saying what to tell. Its design space covers eight kinds of gaps (who is in the scene, how many, what they are like, what they are doing, where, when, how they relate, and what lies beyond the picture, such as speech and thoughts), instantiated through 20 parameterized animations grounded in classical animation principles. A real-time pipeline detects discrepancies between utterance and scene, then selects and parameterizes an animation from the scene's narrative potential and the child's history. Adults interpreted most animations without instruction ($N=120$); children ($N=12$) resolved 46\% of the gaps the system identified with animations, against 13\% without.

\end{abstract}
\begin{teaserfigure}
    \centering
    \includegraphics[width=1\linewidth]{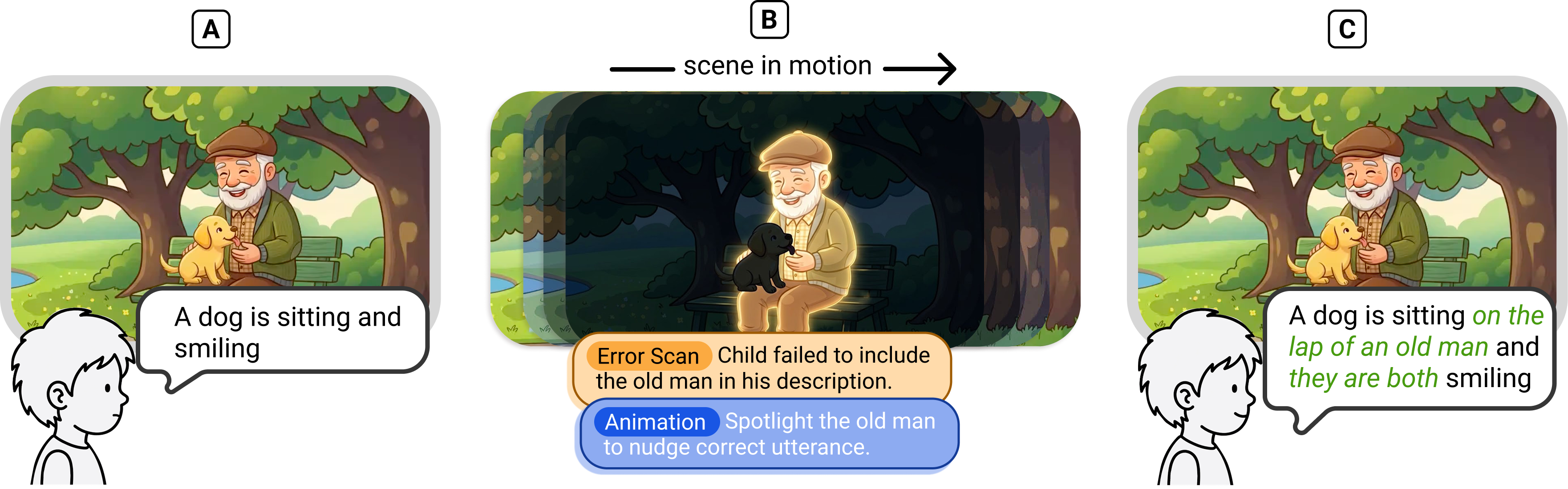}
    \caption{A \system{} interaction cycle. (A) The child narrates an illustrated scene: ``A dog is sitting and smiling.'' (B) Nothing is wrong, but the child has not mentioned the old man the dog sits with: a narrative gap of the kind \emph{who?}, a missing Character. \system selects Spotlight and dims the scene around the man, who glows. (C) The child looks again and says, in their own words, ``A dog is sitting on the lap of an old man and they are both smiling.''}
    \Description{Three panels labeled A, B, and C. Panel A: an illustrated park scene in which an old man with a white beard and a flat cap sits on a bench under a tree with a golden puppy on his lap, both smiling; a line drawing of a child says, A dog is sitting and smiling. Panel B, labeled scene in motion: a stack of frames of the same scene in which everything is darkened except the man, who glows with a pale halo; two tags below read Error Scan, child failed to include the old man in his description, and Animation, spotlight the old man to nudge correct utterance. Panel C: the scene restored to full color; the child says, A dog is sitting on the lap of an old man and they are both smiling, with the added words in green italics.}
    \label{fig:teaser}
\end{teaserfigure}
\maketitle





\input{01_introduction}
\input{02_relatedwork}
\input{03_grammar}

\input{04_ux}

\input{05_system}
\input{06_evaluation}
\input{07_discussion}
\input{08_conclusion}


\bibliographystyle{ACM-Reference-Format}
\bibliography{99_refs}



\end{document}

%% file: 01_introduction.tex
\section{Introduction}

Illustrated scenes, such as the one shown in Figure~\ref{fig:teaser}, are widely used in clinical and educational settings to elicit narratives from children. A speech-language pathologist (SLP) or educator presents an illustration and asks the child to tell the story it depicts. As the child narrates, the expressiveness of the narrative (the characters the child introduces, the actions and settings they describe, the emotions they attribute, and the causal relationships they express) provides insight into their language development~\cite{botting2002, mccabe1991, hedberg1993}. Children with language difficulties often leave much of this information unexpressed. Shown a scene of a small boy running toward a balloon seller in a sunny park, for example, a child may say only \textit{``there is a boy and balloons.''} The picture shows the running, the seller, the setting, and the boy's excitement; the narration does not. We call such a shortfall between what a scene shows and what the child says about it a \emph{narrative gap}. A gap can be an omission, as in this example, or a discrepancy, as when the child says that the boy is walking. A trained educator notices these gaps as the child narrates and scaffolds the storytelling accordingly: they might direct attention to an overlooked action, ask \textit{``What is the boy doing?''}, or probe an unexpressed emotional state with \textit{``How do you think he feels?''} Such interventions help children elaborate narrative elements they may not yet produce spontaneously~\cite{gillam2016skill}.

Narrative gaps are not only a clinical concern. The ability to tell a coherent story from pictures develops throughout the elementary years and predicts later reading comprehension and academic achievement~\cite{catts2002, griffin2004, fey2004}, and most of children's picture-based storytelling happens at home or in the classroom, where no one is trained to recognize each gap and scaffold it~\cite{petersen2011}. Recent AI-driven storytelling tools extend some of this support by acting as a conversational partner that asks the child questions about the story. Yet there is a \textit{tension} in how explicitly such systems should guide the child. A question such as \textit{``What is the boy doing?''} elicits the missing action, but it also \textit{narrows} what the child attends to, presupposes the form of the answer, and hands the initiative to the system. Prior work has shown that direct pedagogical instruction led children to explore a novel object more narrowly and discover fewer of its possibilities than children who encountered it without such explicit instruction~\cite{bonawitz2011double}. This concern is particularly relevant for at-home practice, where the goal is not to provide continued prompting but to help children become increasingly independent storytellers; narrative-intervention practice emphasizes providing only as much support as needed and gradually fading scaffolds as children develop greater proficiency~\cite{spencer2020narrative}. This motivates the central question for our research: \textit{Can we draw a child's attention to a narrative gap without biasing what they say?}

We explore this question through \textit{\system}, an approach that embeds just-in-time visual scaffolding, in the form of animations, directly into the illustrated scenes of a story. The stories are given: each consists of a few illustrated panels, and the child's task is to put each panel into words. As shown in Figure~\ref{fig:teaser}, if a child says only \textit{``there is a boy and balloons''} while the panel shows the boy running toward a balloon seller, \system can animate the boy with motion lines that draw attention to his action, without asking \textit{``What is the boy doing?''} or supplying the word \textit{``running.''} One may object that the picture already tells the child what to say. The picture does show the running, but the child has to notice it, judge it worth telling, and find words for it (running, racing, hurrying after the balloons), and these are precisely the steps that children with language difficulties skip. The animation takes on the first two steps and leaves the third to the child: it shows where to look and what kind of thing is worth telling there (an action, a feeling, a place), without supplying the words. \system turns this mechanism into a scaffold with two layers: the animation raises the salience of the overlooked or misdescribed detail, which draws the child's gaze to it, and its form carries the question to answer there, motion lines asking what the boy is doing where tear particles would ask how he feels; the child answers in their own words.

At the core of \system is a design space that organizes narrative gaps into eight kinds, phrased as the questions a child's narration may leave unanswered: \emph{who} is in the scene (Identity), \emph{how many} there are (Count), \emph{what they are like} (Property), \emph{what they are doing} (Action), \emph{where} they are (Space), \emph{when} events happen (Time), \emph{how} entities and events are \emph{connected} (Relation), and \emph{what lies beyond the picture}, such as speech, thoughts, and intentions (Discourse). Each kind of gap can arise in two forms: as a \emph{discrepancy}, when the child says something the scene contradicts, or as an \emph{opportunity}, when the scene affords a story element the child has not yet expressed. For each kind of gap, the design space provides two to four animations, 20 in total, that make the missing or misdescribed information perceptible: motion lines that exaggerate an action, a spotlight that isolates a character, heart or tear particles that externalize an emotion, a speech bubble that invites dialogue. The animations are grounded in classical animation principles and in visual conventions from cartoons and comics, so that children can interpret them without instruction. A generic highlight would say only ``here.'' Each of the 20 animations also says what kind of gap is here, so that motion lines ask what an entity is doing, a spotlight who it is, a timelapse when, and tear particles how it feels, and none of them says the answer. The design space is a vocabulary of visual questions, the questions a clinician would ask, put into the picture rather than into words. At runtime, a hybrid pipeline analyzes each utterance for both discrepancies and opportunities, selects an animation for the kind of gap it finds, and adapts the animation's intensity to the scene and the child's interaction history. The pipeline completes a full cycle in a median of 5.2 seconds, fast enough for the system to respond before the child's attention shifts. We evaluate \system in three stages, validating the pipeline's intervention choices, testing whether the animations can be interpreted without instruction, and conducting a within-subjects study with 12 children. Children addressed 46\% of the gaps the system identified when animations were present, against 13\% without, and did so in their own words, suggesting that attentional animation can provide a promising \textit{complement} to verbal prompting for child-led storytelling.

Our key contributions include (1) a \textbf{design space} that characterizes children's narrative gaps as eight kinds of unanswered questions and provides 20 attentional animations to make them visible, each animation targeting one kind of gap in both its discrepancy and opportunity forms; (2) an \textbf{end-to-end system} that detects narrative gaps in a child's narration and selects and parameterizes an animation for them in real time; and (3) \textbf{empirical evidence} that \systems animations are largely interpretable without instruction and lead children to address more of the gaps identified by the system, in their own words.

%% file: 02_relatedwork.tex
\section{Related Work}

Oral narratives are widely used to assess children's language development~\cite{botting2002, mccabe1991, hedberg1993}. Producing a story from a picture requires children to identify characters, describe actions, establish spatial and temporal relations, attribute internal states, and connect events causally. The expressiveness and coherence of these narratives are predictive of later reading comprehension and academic achievement~\cite{catts2002, griffin2004, fey2004}. These abilities develop substantially between ages 5 and 12, progressing from labeling and listing toward increasingly coherent, causally structured, multi-episode narratives, with an important transition around ages 7 to 9~\cite{berman2009, peterson1983, stein1979, stadler2005, ukrainetz2005}. This developmental trajectory motivates our focus on children beginning at age seven, when they can produce connected narratives but are still consolidating many of the narrative skills that \system seeks to scaffold. Further, children with developmental language disorder (DLD) and related language difficulties often face particular challenges in this process. Their stories tend to be shorter, less cohesive, and structurally simpler than those of typically developing peers~\cite{leonard2014, botting2002, greenhalgh2001, liles1995, liles1993, vandewalle2012}. They may omit initiating events, internal responses, and causal connectives, and struggle with the expressive elaboration that gives narratives their ``art''~\cite{merritt1989, ukrainetz2009, epstein2009}. 


These observations motivate a long tradition of narrative intervention and assessment, and they frame the tension introduced above: children benefit from targeted scaffolding, but that scaffolding should not supply the missing language or take over the storytelling. We draw on three bodies of prior work: narrative scaffolding and assessment, interactive technologies for children's storytelling, and animation as attentional communication.

\subsection{Narrative Scaffolding and Assessment}
\label{sec:intervention}
\label{sec:storytelling}

Structured narrative interventions teach story structure and literate language by explicitly scaffolding the elements children have difficulty producing. Approaches such as SKILL~\cite{gillam2016skill, gillam2018} and related interventions~\cite{swanson2005, ukrainetz2006, joffe2019, pauls2021} use modeling, prompting, questioning, and other forms of guided practice to support children's narrative production. Such interventions can yield moderate-to-large gains, although systematic reviews note that the evidence base remains relatively small and methodologically uneven~\cite{petersen2011}. Importantly, these approaches typically depend on a clinician or trained educator to recognize what is missing from a child's narration, decide how to scaffold it, and monitor progress over time.

Narrative assessment frameworks provide a complementary account of \emph{what} proficient storytelling entails. Story grammar analysis, rooted in schema theory~\cite{stein1979, schank1975}, identifies canonical narrative components such as setting, initiating event, internal response, and resolution, while standardized instruments operationalize these constructs for clinical assessment~\cite{gillampearson2004, renfrew1997, schneider2006, westerveld2015}. We build in particular on the \textbf{Monitoring Indicators of Scholarly Language} (MISL) framework~\cite{gillam2017, israelsen2022}, which draws on theories of discourse comprehension~\cite{graesser1997, kintsch1998, kintsch1988, zwaan1998} and distinguishes story-level macrostructure from sentence-level features of literate language. We use this element-level characterization because it provides a concrete vocabulary for identifying opportunities for targeted scaffolding. Recent work further suggests that portions of this assessment process can be automated: machine-learning and LLM-based approaches increasingly approach human reliability in rating narrative macrostructure and literate-language use~\cite{jones2019, fox2022}. \system builds on this possibility by analyzing children's narration during the storytelling interaction rather than only after the story is complete.

\subsection{Interactive Technologies for Children's Storytelling}
\label{sec:storytellingtools}

A large body of HCI research has explored technologies for supporting children's storytelling, ranging from open-ended creation environments~\cite{rubegni2014, mckinley2008} and embodied robots~\cite{korywestlund2017, ligthart2020} to increasingly prevalent generative-AI storytelling systems~\cite{zhang2022storybuddy, fan2025storyprompt, toyteller, Ye:2025, Chen:2025, Panchal:2024}; scoping reviews and surveys document this rapidly growing design space~\cite{im2025review, riedl2013}. Much of this work positions technology as a partner in story \emph{creation}: LLM-driven tools help children and authors generate plots, characters, and illustrations~\cite{Wang:2024, Alvarez:2022, Mishra:2025, simon2022, Bensaid:2021}; comic- and avatar-based systems support collaborative and embodied expression~\cite{rutta2020, zarei2020, zarei2021}; and other systems transform children's drawings or image streams into narrated stories~\cite{zhang2022storydrawer, zhou2025, wang2020}.

The extent to which such systems support language \emph{learning}, however, remains less clear. Mathemyths~\cite{zhang2024mathemyths} found that children co-creating stories with an LLM developed mathematical language at rates comparable to children co-creating with a human partner, whereas Storyfier~\cite{peng2023} increased engagement while resulting in poorer recall of target vocabulary than a non-AI baseline. These findings echo broader work in child-computer interaction showing that engagement, usability, and learning are related but distinct outcomes~\cite{sim2006, tisza2022}, and that children's learning technologies require age-appropriate, longitudinal, and participatory forms of evaluation~\cite{sim2016, sim2022, mcnaney2013, mcknight2013, mann2016, lamichhane2023}.

Most of these systems also support producing content \emph{with} or \emph{for} the child, rather than attending to and scaffolding the child's own ongoing narrative production. Systems closer to clinical practice have begun to take on more targeted scaffolding roles. AACessTalk~\cite{choi2025aacesstalk}, for example, uses context-aware AI guidance and vocabulary-card recommendations to support communication between minimally verbal autistic children and their parents, though its feedback scaffolds the dyad's conversation rather than a child's narrative production. OSOS~\cite{Lee:2024}, which like \system was co-designed with speech-language pathologists, uses generative AI to adapt intervention materials to individual children, but does so by generating personalized storybooks for children to read. In contrast, \system responds contingently to the story a child is currently telling. Rather than generating story content or acting primarily as a conversational partner, it uses visual feedback to scaffold the child's own narrative production in real time.

\subsection{Animation as Attentional Communication}
\label{sec:rwanim}
\label{sec:rwperception}

Animation has long served as a communicative medium in user interfaces: motion can convey state, affordance, and transition beyond what static graphics provide~\cite{baecker1990}, techniques borrowed from cartoons can make interface behavior legible~\cite{chang1993}, and later work extended its roles from toolkit support~\cite{hudson1993animation} to kinetic typography~\cite{lee2002kinetic} and a broad range of communicative uses~\cite{chevalier2016}. The closest precedent is Kineticons~\cite{harrison2011kineticons}, a vocabulary of motion behaviors such as bouncing, shaking, and pulsing that communicate interface states once users have learned the mapping. \system takes the complementary direction: it starts from a semantic objective, a narrative element that requires attention, and selects a visual transformation that makes it perceptible, drawing on conventions children already know from comics and cartoons, such as emanata~\cite{walker1980} and motion lines~\cite{mccloud1993}, and on the classical principles of staging and exaggeration~\cite{thomas1981, lasseter1987}, which direct the viewer's eye to the one element that matters.

Related systems map speech to visual augmentation in real time: RealityTalk~\cite{liao2022realitytalk} lets presenters trigger augmented graphics through speech, Visual Captions~\cite{liu2023visualcaptions} introduces visual suggestions during conversation, and DrawTalking~\cite{rosenberg2024drawtalking} animates users' sketches as they narrate. These systems augment what a speaker is already communicating; \system uses speech-contingent visuals to point at what a narration has misstated or left unexpressed. Generative models make the scenes for such feedback easy to construct and analyze, through image generation~\cite{rombach2022, ramesh2022}, vision-language description~\cite{liu2024visual, wu2023}, prompt design~\cite{oppenlaender2024}, cross-image consistency~\cite{li2025}, and multi-scene composition~\cite{xiao2025}, and they extend to animation itself~\cite{lin2026visuallyrics}. \system uses them to prepare and analyze its scenes, but its feedback animations are procedural templates applied to pre-extracted entities, which keeps them fast, deterministic, and controllable.

Adding motion is not enough, however. Animated graphics do not necessarily outperform informationally equivalent static ones, and Tversky et al.~\cite{tversky2002} proposed two principles for when they do: the \emph{Congruence Principle}, whereby the form of the visual change corresponds to the structure of the concept being communicated, and the \emph{Apprehension Principle}, whereby the change remains simple and slow enough to be perceived accurately. Animations that violate them add extraneous cognitive load~\cite{klepsch2017, zu2021}; staged, predictable transitions help viewers track change~\cite{heer2007}, whereas added motion complexity can impair it~\cite{chevalier2014}. Interpretability also depends on shared conventions: Kineticons found substantial variation in how reliably different motions conveyed their meaning~\cite{harrison2011kineticons}, and expressive animated delivery affects children's learning and retention~\cite{korywestlund2017}. We therefore treat Congruence and Apprehension as design constraints: \systems animations are localized, short, and single-purpose, congruent with the gap they point at, and built from conventions children already know.

%% file: 03_grammar.tex
\section{\system Design Space}
\label{sec:designspace}

\system aims to translate a narrative gap into a visual change that draws the child's attention without supplying the language they should produce. Its design space answers three questions: \emph{what kind of gap} the narration leaves (eight kinds, Section~\ref{sec:gaps}), \emph{in which form} the gap arises (a discrepancy or an unexpressed opportunity, Section~\ref{sec:gap-forms}), and \emph{how} the gap can be made perceptible in the scene (20 animations, Section~\ref{sec:animations}).

Four requirements guided the design space. They emerged from working closely with a speech-language pathologist (SLP), who is also an author of this paper, on how visual feedback could support children's narrative production without replacing the reasoning and language production that storytelling practice is intended to elicit; from prior work on perceptually effective animation~\cite{tversky2002, thomas1981, lasseter1987}; and from the constraints of manipulating illustrated scenes in real time.

Concretely, \system feedback should provide \textbf{implicit scaffolding}, making a narrative gap salient without prescribing what the child should say; be \textbf{legible without instruction}, avoiding arbitrary visual mappings that must be learned; be \textbf{locally addressable}, so that it can target a particular entity, relation, group, or scene; and be \textbf{parameterizable}, allowing its visual salience to be adapted to the scene and interaction context. These requirements constrain each layer of the design space described below.

\subsection{Eight Kinds of Narrative Gaps}
\label{sec:gaps}

A narrative gap is a shortfall between what a scene shows and what the child says about it. To organize gaps in a way that visual feedback can act upon, we grouped them by the information a child would need to notice in order to close them, rather than by the linguistic form of the shortfall. Taxonomies of children's oral-language errors~\cite{dulay1982, paul2012} classify that form (an omission, a wrong tense, an ambiguous pronoun), but a form label does not say what should change in the picture. An ambiguous pronoun, a vague reference, and a confusion between two characters are three different errors that all require the same thing of the child: to look at which entity is meant. A tense inconsistency and a temporal-ordering mistake both require attention to when an event occurs. Grouping gaps by what must be noticed, and merging groups that the same family of visual strategies could address, yielded eight kinds of gaps. We phrase each as the question that the child's narration leaves unanswered. Together they progress from concrete information about individual entities toward abstract relations between events:

\begin{itemize}
    \item \textbf{Who? (Identity)} The child omits, confuses, or refers ambiguously to a character or object, or could introduce a character more explicitly. Closing the gap requires looking at \emph{which entity} is meant.

    \item \textbf{How many? (Count)} The child mentions an entity that is not in the scene, omits one of several similar entities, or refers ambiguously among them. Closing the gap requires noticing \emph{which entities are present}, and in what number.

    \item \textbf{What are they like? (Property)} A visible attribute such as color, size, or texture, or an internal or physical state that can be externalized, such as sadness, affection, heat, or cold. Closing the gap requires looking closely at the entity itself.

    \item \textbf{What are they doing? (Action)} An action that is overlooked or described incorrectly. Closing the gap requires noticing the movement or activity depicted.

    \item \textbf{Where? (Space)} Position, containment, occlusion, and the physical organization of the scene. Closing the gap requires noticing where an entity is, including behind another one.

    \item \textbf{When? (Time)} Temporal progression, duration, and the ordering of events. Closing the gap requires noticing that time passes within and between scenes.

    \item \textbf{How are they connected? (Relation)} Grouping, separation, and causal links between entities or events, which cannot be conveyed by attending to a single entity in isolation.

    \item \textbf{What lies beyond the picture? (Discourse)} Dialogue, thoughts, intentions, and narrative importance, which no static illustration shows directly, as well as the form of the child's own utterance.
\end{itemize}

Gaps that the illustration cannot address, such as errors of pronunciation, fall outside the scope of \system.

\subsection{Two Forms of Gap: Discrepancies and Opportunities}
\label{sec:gap-forms}

\begin{table*}[t]
\centering
\caption{The 15 elements of the MISL rubric~\cite{gillam2017, israelsen2022}. Macrostructure elements describe story-level components, while microstructure elements describe sentence-level devices of literate language. \system uses these elements to characterize narrative opportunities that can be scaffolded through the design space.}
\label{tab:misl}
\small
\renewcommand{\arraystretch}{1.15}
\begin{tabular}{@{}clll@{}}
\toprule
& \textbf{Element} & \textbf{What it captures} & \textbf{Example} \\
\midrule
\multicolumn{4}{@{}l}{\textit{Macrostructure}} \\
\macrotag{CH} & Character & Naming, specifying, and distinguishing characters & ``a shy little boy'' \\
\macrotag{S} & Setting & Locating the story in time and place & ``one sunny day, in the park'' \\
\macrotag{IE} & Initiating Event & The problem that sets the story in motion & ``his balloon flew away'' \\
\macrotag{IR} & Internal Response & How characters feel about the initiating event & ``the boy was very sad'' \\
\macrotag{P} & Plan & Characters' intentions to act on the initiating event & ``he decided to get it back'' \\
\macrotag{A} & Action & Actions taken in response to the initiating event & ``he ran after the balloon'' \\
\macrotag{CO} & Consequence & The outcome tied to the initiating event & ``at last, he caught it'' \\
\midrule
\multicolumn{4}{@{}l}{\textit{Microstructure}} \\
\microtag{CC} & Coordinating Conjunctions & Linking clauses of equal weight & \textit{for, and, nor, but, or, yet, so} \\
\microtag{SC} & Subordinating Conjunctions & Linking a dependent clause to a main clause & \textit{because, when, after, while, if} \\
\microtag{M} & Mental Verbs & Verbs of thought and intention & \textit{thought, decided, wanted} \\
\microtag{L} & Linguistic Verbs & Verbs of speech & \textit{said, told, yelled} \\
\microtag{ADV} & Adverbs & Modifying manner, time, or degree & \textit{slowly, suddenly, very} \\
\microtag{ENP} & Elaborated Noun Phrases & Noun phrases carrying two or more modifiers & ``the big red dog'' \\
\microtag{G} & Grammaticality & Freedom from morphological and syntactic errors & ``he runned'' $\rightarrow$ ``he ran'' \\
\microtag{T} & Tense & Consistency of verb tense across the narrative & ``he runs\ldots{} he caught'' \\
\bottomrule
\end{tabular}
\end{table*}

Each kind of gap can arise in two forms, and the eight kinds are not tied to either. \emph{Action} can foreground an action the child described incorrectly or one the child has simply not mentioned; \emph{Property} can surface a wrong description of an object or invite elaboration of a character's emotional state.

A \emph{discrepancy} is a clear conflict between the child's utterance and the illustrated scene, or with the child's own immediately preceding utterance. The correction space is intentionally tolerant: visual feedback is intended for clear discrepancies that can be grounded in the scene, not for every departure from an idealized narrative, so plausible interpretations, creative elaborations, and minor age-appropriate variations remain untouched.

An \emph{opportunity} arises when the narration is entirely consistent with the illustration but leaves story content unexpressed: a child may identify a character and describe an action without mentioning the character's internal response, the causal relationship between events, or descriptive details that would make the narrative richer. To characterize these opportunities, we use the 15 elements of the Monitoring Indicators of Scholarly Language (MISL) rubric~\cite{gillam2017, israelsen2022}, which distinguishes seven macrostructure elements that capture story-level organization from eight microstructure elements of literate language (Table~\ref{tab:misl}). These elements give \system a vocabulary for describing \emph{what kind of narrative content} a scene affords and what the child has or has not yet expressed. We do not treat a missing element as an error but as an \emph{enrichment objective}, which lets \system separate correction from enrichment while representing both as something the feedback should help the child notice.

The mapping between MISL elements and the eight kinds of gaps is many-to-many. Internal Response, for example, may be addressed as a \emph{Property} gap when an emotion can be externalized visually, or as a \emph{Discourse} gap when a thought or intention is better represented symbolically. Conversely, a \emph{Property} cue may support Internal Response in one context and an Elaborated Noun Phrase or Adverb in another. For a discrepancy, \system moves from the detected conflict to the kind of gap it implicates. For an opportunity, it starts from a MISL element that the scene affords and determines which kind of gap, once made salient, would invite that content. In both cases, the kind of gap defines \emph{what the feedback should direct the child's attention toward}, while leaving open how that goal should be rendered.

\subsection{Twenty Animations}
\label{sec:animations}

\begin{figure*}
    \centering
    \includegraphics[width=\textwidth]{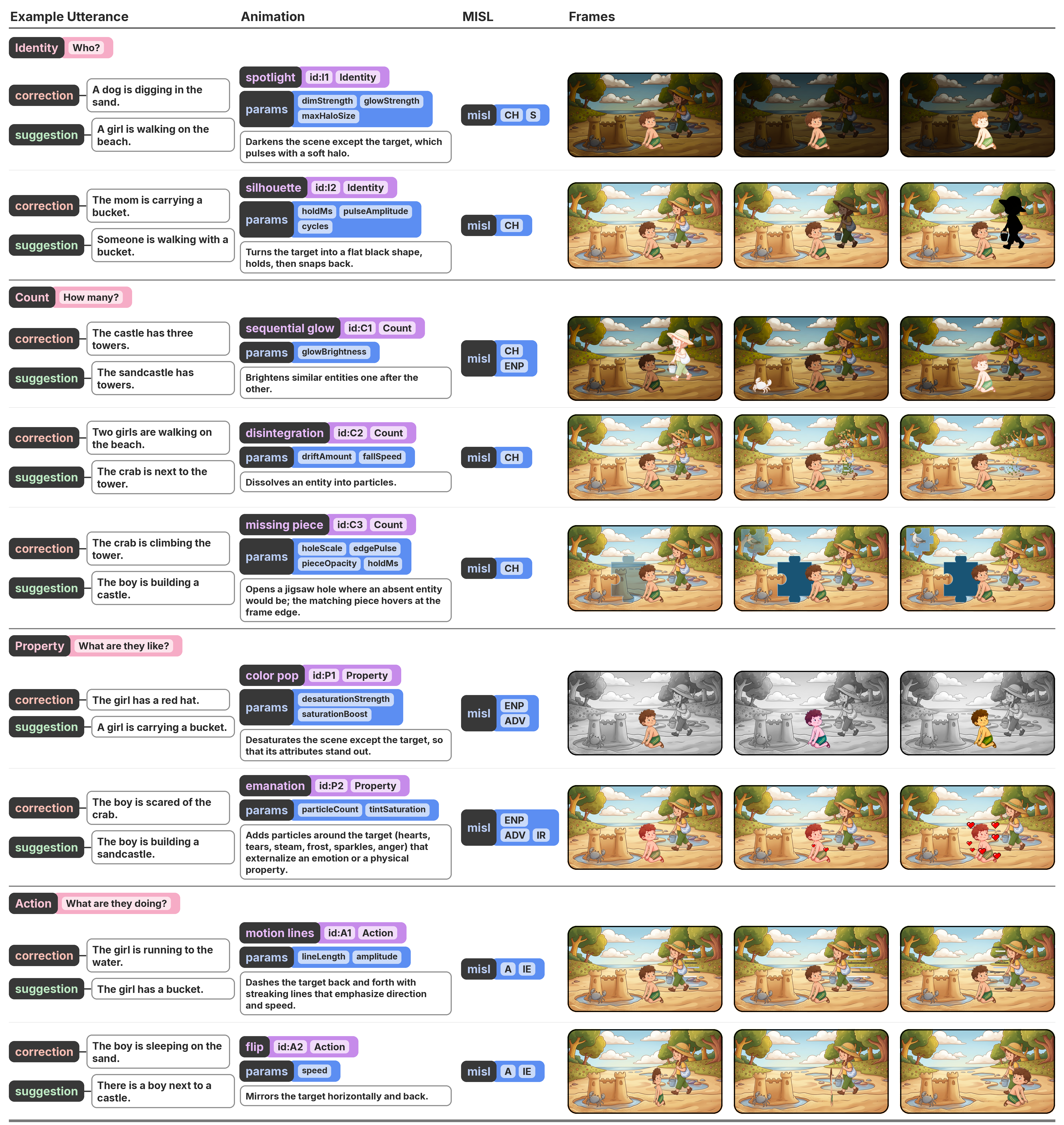}
    \caption{The design space, part 1: Identity (Who?), Count (How many?), Property (What are they like?), and Action (What are they doing?), nine of the twenty animations. For each animation, from left to right: an example utterance containing an error the animation corrects and one leaving an opportunity the animation invites the child to take; its name, identifier, and category, the parameters that salience levels adjust, and its visual behavior; the MISL elements it can address; and three frames of the animation on the practice scene. Emanation (P2) is one animation with six particle variants.}
\Description{A table with four columns headed Example Utterance, Animation, MISL, and Frames, divided into four bands by category headers reading Identity, Who?; Count, How many?; Property, What are they like?; and Action, What are they doing?. Each row is one animation. The first column holds two rounded boxes labeled correction and suggestion, each with a short sentence such as A dog is digging in the sand. The second column shows the animation name in a purple tag with its identifier and category, a blue tag listing parameter names such as dimStrength or holdMs, and a sentence describing the visual behavior. The third column shows dark tags with MISL codes such as CH, S, ENP, ADV, IR, A, or IE. The fourth column shows three stills of the same cartoon beach scene, a boy kneeling by a sandcastle, a girl in a sun hat carrying a bucket, and a crab, at the beginning, middle, and end of the animation. The nine rows are Spotlight (I1), which darkens the scene except the target, which pulses with a halo; Silhouette (I2), which turns the target into a flat black shape, holds, then snaps back; Sequential Glow (C1), which brightens similar entities one after the other; Disintegration (C2), which dissolves an entity into particles; Missing Piece (C3), which opens a jigsaw hole where an absent entity would be while the matching piece hovers at the frame edge; Color Pop (P1), which desaturates the scene except the target; Emanation (P2), which adds particles such as hearts or tears around the target; Motion Lines (A1), which dashes the target back and forth with streaking lines; and Flip (A2), which mirrors the target horizontally and back.}
    \label{fig:animations-1}
\end{figure*}

\begin{figure*}
    \centering
    \includegraphics[width=1\textwidth]{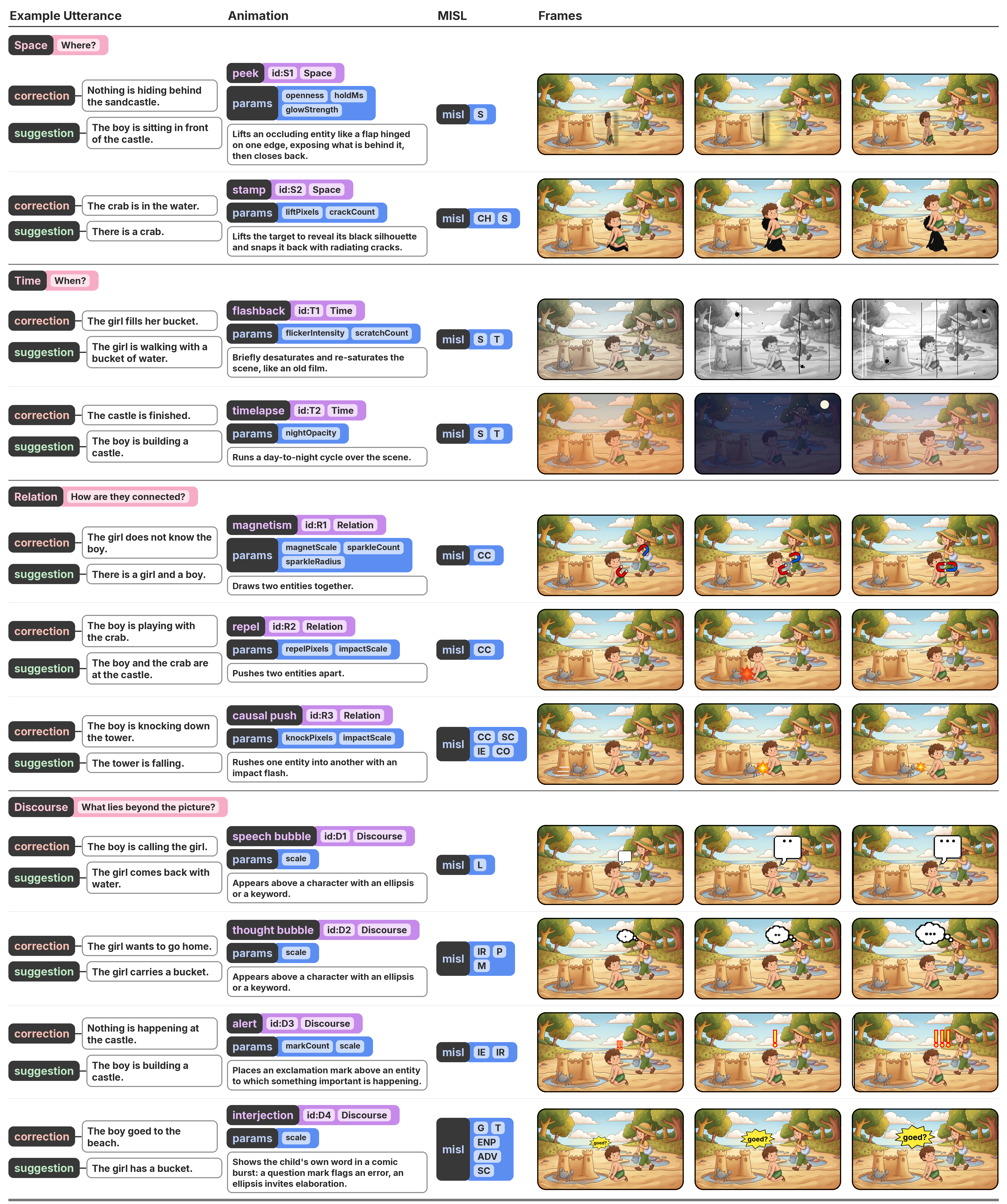}
    \caption{The design space, part 2: Space (Where?), Time (When?), Relation (How are they connected?), and Discourse (What lies beyond the picture?), the eleven remaining animations. Same columns as Figure~\ref{fig:animations-1}: example utterances for a correction and a suggestion; name, identifier, category, parameters, and visual behavior; MISL elements; and three frames on the practice scene.}
\Description{A table with the same four columns as the previous one, Example Utterance, Animation, MISL, and Frames, divided into four bands by category headers reading Space, Where?; Time, When?; Relation, How are they connected?; and Discourse, What lies beyond the picture?. Each row is one animation, with two example sentences labeled correction and suggestion, the animation name in a purple tag with its identifier and category, a blue tag of parameter names, a sentence describing the visual behavior, dark tags with MISL codes, and three stills of the cartoon beach scene with the boy, the sandcastle, the girl with a bucket, and the crab. The eleven rows are Peek (S1), which lifts an occluding entity like a flap hinged on one edge, exposing what is behind it, then closes back; Stamp (S2), which lifts the target to reveal its black silhouette and snaps it back with radiating cracks; Flashback (T1), which briefly desaturates and re-saturates the scene like an old film, with scratches; Timelapse (T2), which runs a day-to-night cycle over the scene; Magnetism (R1), which draws two entities together with magnet icons; Repel (R2), which pushes two entities apart; Causal Push (R3), which rushes one entity into another with an impact flash; Speech Bubble (D1) and Thought Bubble (D2), which appear above a character with an ellipsis or a keyword; Alert (D3), which places exclamation marks above an entity to which something important is happening; and Interjection (D4), which shows the child's own word in a comic burst, with a question mark for an error and an ellipsis to invite elaboration.}
    \label{fig:animations-2}
\end{figure*}

\begin{figure*}[t]
    \centering
    \includegraphics[width=0.6\linewidth]{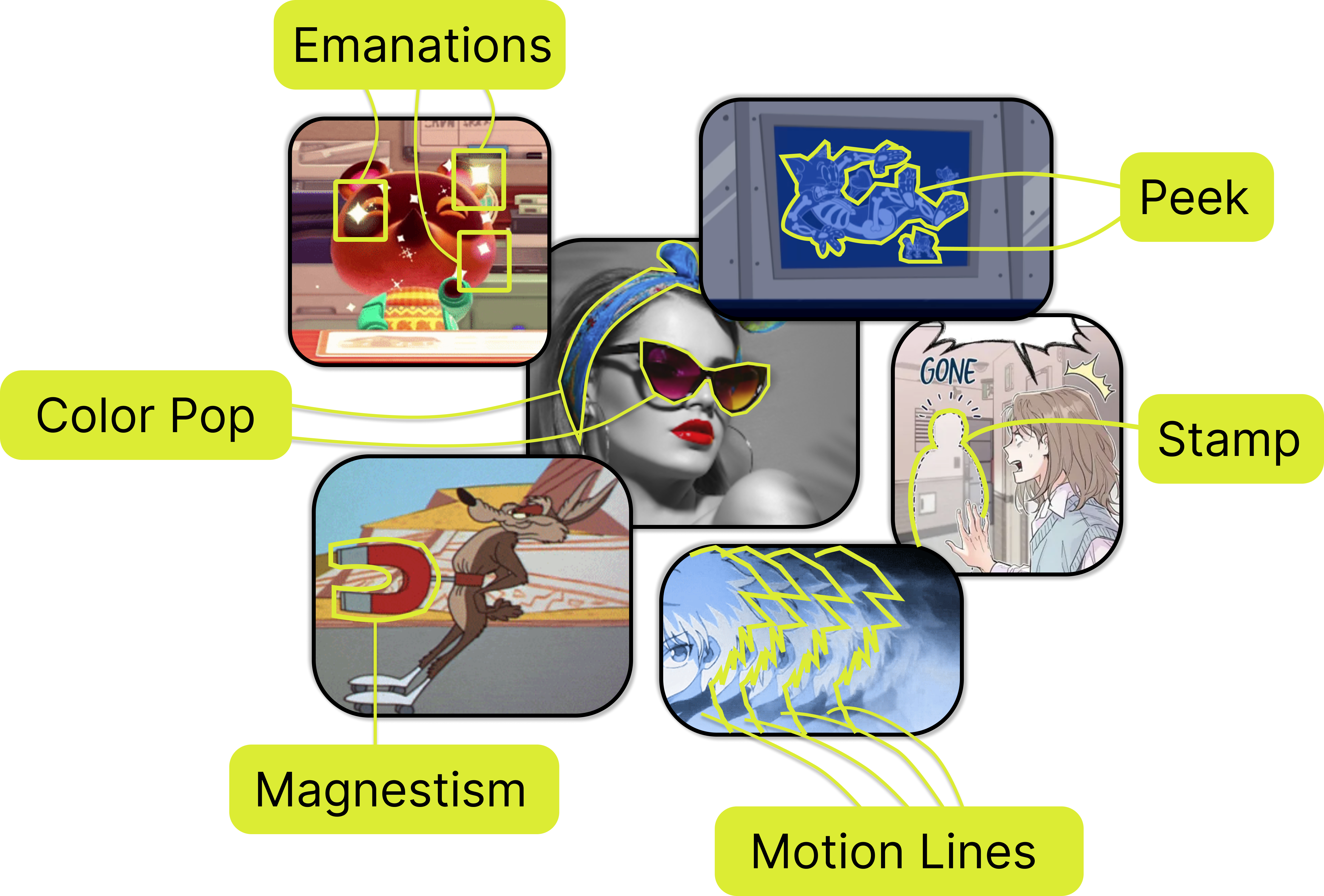}
    \caption{Six conventions of animated cartoons that the design space borrows, each labeled with the animation it inspired: symbols drawn around a character (Emanation), a screen revealing what a container hides (Peek), a single colored element in a grayscale frame (Color Pop), a character-shaped hole left in a wall (Stamp), a horseshoe magnet pulling a character (Magnetism), and speed lines with repeated afterimages (Motion Lines). Stills from commercial animated films and series.}
    \Description{Six rounded stills from animated cartoons arranged in a loose cluster, each connected by a thin line to a yellow label. Top left, labeled Emanation: small sparkles and symbols drawn around a character. Top right, labeled Peek: a blue screen showing the outline of what is hidden inside a container. Left, labeled Color Pop: a grayscale drawing of a woman in which only her red lips and her sunglasses are in color. Right, labeled Stamp: a character-shaped hole in a wall next to the word GONE, left by a character who ran through it. Bottom left, labeled Magnetism: a coyote-like character in a desert holding a large horseshoe magnet. Bottom right, labeled Motion Lines: a character drawn with repeated afterimages and jagged speed lines. Yellow outlines on each still trace the element the label refers to.}
    \label{fig:inspirations}
\end{figure*}

The last question is \emph{how} a gap is made perceptible. Rather than inventing an arbitrary visual code, we drew on conventions that illustrated and animated media already use to direct attention or communicate without dialogue: McCloud's account of the visual language of comics~\cite{mccloud1993}, Walker's catalog of emanata~\cite{walker1980}, and the classical principles of staging, exaggeration, anticipation, and motion~\cite{thomas1981, lasseter1987}. Figure~\ref{fig:inspirations} shows six of these conventions as cartoons use them, next to the animation each one became. Following the requirements above, we retained conventions that have a constrained interpretation when viewed in context, generalize across characters and objects, can be implemented by programmatically manipulating scene layers rather than character-specific geometry, and can be parameterized. The result is 20 animations, each assigned to one kind of gap (Figures~\ref{fig:animations-1} and~\ref{fig:animations-2}).

\begin{itemize}
    \item \textbf{Who?} \emph{Spotlight} darkens the scene except the target, which pulses with a soft halo. \emph{Silhouette} turns the target into a flat black shape, holds, then snaps back, asking who or what it is.

    \item \textbf{How many?} \emph{Sequential Glow} brightens similar entities one after the other. \emph{Disintegration} dissolves a wrongly mentioned entity into particles. \emph{Missing Piece} opens a jigsaw-shaped hole where an absent entity would be, while the matching piece hovers at the edge of the frame.

    \item \textbf{What are they like?} \emph{Color Pop} desaturates the scene except the target, so that its color, size, and texture stand out. \emph{Emanation} adds particles around the target (hearts, tears, steam, frost, and other conventional signs) that externalize an emotion or a physical property.

    \item \textbf{What are they doing?} \emph{Motion Lines} dashes the target back and forth with streaking lines that emphasize direction and speed. \emph{Flip} mirrors the target horizontally and back, signaling that its action has not been described or is unclear.

    \item \textbf{Where?} \emph{Peek} lifts an occluding entity like a flap hinged on one edge, exposing what is behind it. \emph{Stamp} lifts the target to reveal its black silhouette and snaps it back with radiating cracks, anchoring it here rather than elsewhere.

    \item \textbf{When?} \emph{Flashback} briefly desaturates and re-saturates the target, signaling that an event has already occurred. \emph{Timelapse} runs a day-to-night cycle over the scene.

    \item \textbf{How are they connected?} \emph{Magnetism} draws two entities together. \emph{Repel} pushes them apart. \emph{Causal Push} rushes one entity into another with an impact flash, visualizing that one event causes the next.

    \item \textbf{What lies beyond the picture?} \emph{Speech Bubble} and \emph{Thought Bubble} appear above a character with an ellipsis or a keyword. \emph{Alert} places an exclamation mark above an entity to which something important is happening. \emph{Interjection} shows the child's own problematic word in a comic-style burst with a question mark, making a speech error visible for self-correction.
\end{itemize}

Silhouette, Missing Piece, and Peek replace three animations that our interpretability study found hard to interpret; we describe the original designs and the redesign in Section~\ref{sec:eval-interpret}.

The animations use two broad mechanisms. The first \emph{transforms information already present in the scene}: Spotlight isolates an entity by dimming its surroundings, Color Pop foregrounds its visual attributes through saturation, Motion Lines exaggerates the direction and speed of an action, Peek lifts an occluder, and the temporal and relational animations move visible entities to show how they are situated in time or with respect to one another. These animations redistribute perceptual salience: the information is already in the illustration but becomes harder to overlook.

The second \emph{adds a temporary visual representation} for information the static illustration does not show: Emanation externalizes an emotion or physical property through particles, Speech Bubble and Thought Bubble make dialogue, thoughts, or intentions available as possibilities, Alert marks an entity as narratively important, Interjection shows part of the child's own utterance for reconsideration, and Missing Piece stands for an entity absent from the scene but relevant to the story.

\begin{figure*}[t]
    \centering
    \includegraphics[width=0.6\textwidth]{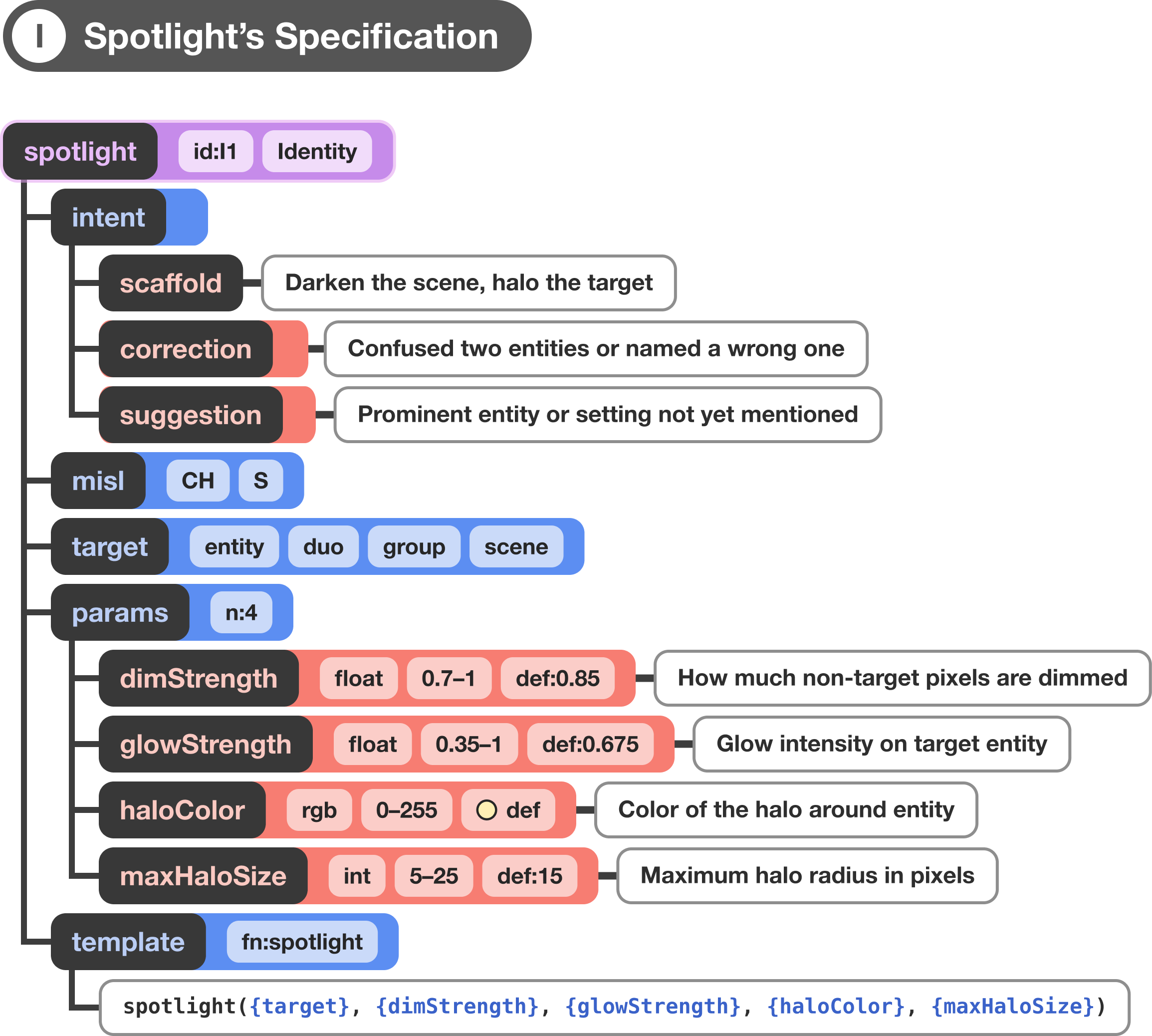}
    \caption{Specification of Spotlight (I1), the first animation of the Identity kind. The specification records its scaffolding behavior and the situations in which the correction and enrichment paths may select it, the MISL elements it can address (Character and Setting), the kinds of target it accepts (a single entity, a duo, a group, or the whole scene), and its four parameters with their type, range, and default; each numeric default sits at the middle of its range, the starting point from which salience levels move it. The rendering template maps the target and the parameter values to the function that runs the animation on the layered scene.}
    \Description{Tree diagram of the Spotlight specification. The root node reads spotlight, id I1, Identity. Five child nodes follow. Intent has three leaves: scaffold, darken the scene and halo the target; correction, confused two entities or named a wrong one; suggestion, prominent entity or setting not yet mentioned. MISL lists CH and S. Target lists entity, duo, group, and scene. Params lists four parameters: dimStrength, a float from 0.7 to 1 with default 0.85, how much non-target pixels are dimmed; glowStrength, a float from 0.35 to 1 with default 0.675, glow intensity on the target entity; haloColor, an RGB value from 0 to 255 with a pale yellow default, color of the halo around the entity; maxHaloSize, an integer from 5 to 25 with default 15, maximum halo radius in pixels. Template reads fn:spotlight and expands to the call spotlight(target, dimStrength, glowStrength, haloColor, maxHaloSize).}
    \label{fig:spec}
\end{figure*}

Each animation is represented as a \emph{parameterized animation specification} (Figure~\ref{fig:spec}) that records the kind of gap it targets, the kinds of scene targets to which it can be applied, its correction and enrichment intents, and a set of parameters that control its visual realization. A corresponding rendering template defines how those parameters transform scene elements or generate temporary overlays. Across the 20 animations, the current implementation exposes 47 adjustable parameters, continuous values such as the amplitude and speed of Motion Lines or the saturation of Color Pop, and discrete choices such as the particle type of Emanation or the content of a bubble.

Parameterization separates the \emph{semantic intent} of a scaffolding act from its \emph{visual intensity}: the same animation can be shown subtly when an opportunity first arises, more saliently when a similar gap previously went unaddressed, and more faintly once the child produces the element on their own, without changing the kind of gap it targets. Section~\ref{sec:architecture} describes how \system makes these selections and adapts their parameters during interaction.

%% file: 04_ux.tex
\section{User Experience}
\label{sec:ux}

\begin{figure*}[t]
    \centering
    \includegraphics[width=\textwidth]{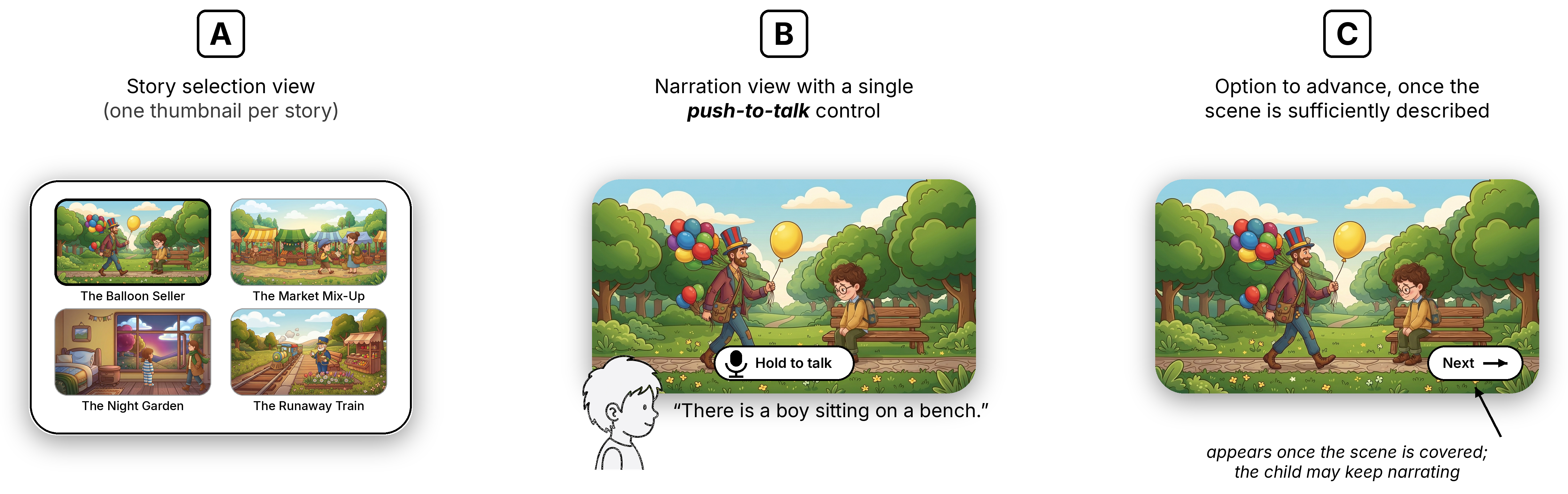}
    \caption{The \system{} interface. (A) The story selection view shows one thumbnail per story. (B) The narration view shows one illustrated panel at a time, with a single push-to-talk control. (C) Once the scene has been sufficiently described, the option to advance appears; the child can move on or keep narrating.}
    \Description{Three panels labeled A, B, and C. Panel A, story selection view: a rounded window with four story thumbnails in a two-by-two grid, The Balloon Seller, The Market Mix-Up, The Night Garden, and The Runaway Train, the first one outlined. Panel B, narration view: the first scene of The Balloon Seller fills a rounded frame, a balloon seller carrying a bunch of colorful balloons and holding out a yellow one toward a boy sitting on a park bench, with a single button reading Hold to talk and a microphone icon at the bottom center; a line drawing of a child faces the scene and says, There is a boy sitting on a bench. Panel C, option to advance: the same scene with a button reading Next and an arrow at the bottom right, annotated as appearing once the scene is covered while the child may keep narrating.}
    \label{fig:ux-setup}
\end{figure*}

Having described what \system can make visible and how, we now turn to how children experience it. \system is implemented as a web-based application and is intentionally designed around a \textit{minimal} scene-centered interaction. Children first choose a story from a selection view showing a thumbnail of its opening scene, then enter a narration view that presents one illustrated panel at a time. In the narration view, the illustration occupies most of the screen, and the child narrates using a single push-to-talk control (Figure~\ref{fig:ux-setup}). Stories typically consist of five to eight illustrated scene panels. The option to advance to the next panel appears once \system determines that the child's narration has sufficiently covered the current scene, allowing the interaction to progress without requiring the child to navigate menus or make explicit interface decisions. To better understand \systems scaffolding, let us follow Alex, a seven year old who is practicing picture based storytelling.

\subsection{Setup}
On first use, Alex is shown a short onboarding screen that explains how to interact with the system. The instructions encourage him to describe not only what is happening in each scene, but also what the characters see, feel, think, or say. Alex then completes two short practice stories that introduce the push-to-talk interaction and let him experience the scene responding visually before beginning a longer story. During this practice, he is told that the picture may move or change after he speaks and is encouraged to look again and continue the story in his own words. The system does not explain the meaning of individual animations; the practice stories are intended only to familiarize children with \system. 

Alex selects \textit{The Balloon Seller} and sees its first illustrated panel. He holds the push-to-talk control and begins narrating what he sees. When he finishes, \system transcribes and analyzes the utterance and determines whether there is an opportunity for visual scaffolding. If no intervention is needed, the scene remains unchanged. Otherwise, \system plays one of the 20 animations of Section~\ref{sec:designspace} on a relevant element of the illustration. The animation loops until Alex presses to speak again.

\begin{figure*}[t]
    \centering
    \includegraphics[width=\textwidth]{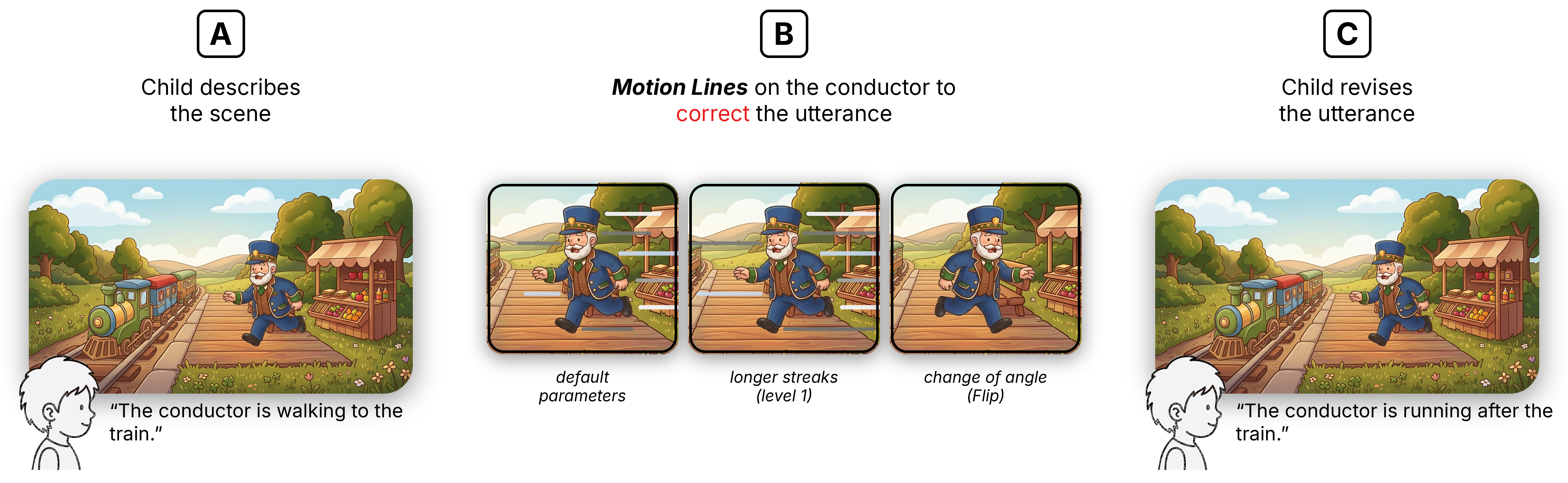}
    \caption{Correcting an error. (A) Alex describes the conductor as walking. (B) \system{} plays Motion Lines on the conductor; if the error persists, the streaks lengthen, then the bursts of movement widen, and once every parameter is at its maximum the same action is shown through a different animation (Flip). (C) Alex revises the description in his own words.}
    \Description{Three panels labeled A, B, and C. Panel A, child describes the scene: a scene from The Runaway Train, a conductor in a blue uniform running along a wooden platform between a small green locomotive on the left and a fruit stall on the right; a line drawing of a child says, The conductor is walking to the train. Panel B, Motion Lines on the conductor to correct the utterance: three square close-ups of the conductor. In the first, labeled default parameters, short white horizontal streaks trail behind him; in the second, labeled longer streaks, level 1, the streaks are longer; in the third, labeled change of angle, Flip, the conductor is mirrored to face the other way and the streaks are gone. Panel C, child revises the utterance: the same scene as panel A; the child says, The conductor is running after the train.}
    \label{fig:ux-correction}
\end{figure*}

\subsection{Error Correction}
In \textit{The Runaway Train}, Alex reaches a scene showing a conductor running after his train (Figure~\ref{fig:ux-correction}). Alex says, \textit{``The conductor is \textbf{walking} to the train.''} The description is largely consistent with the scene, but mischaracterizes the conductor's action: a discrepancy of the kind \emph{what are they doing?} \system responds by animating the conductor with \textbf{Motion Lines}, exaggerating his movement and drawing attention to the speed of the action. Rather than telling Alex that the conductor is running or asking him to try again, the animation gives him an opportunity to reconsider what he sees and revise the description in his own words, for example, \textit{``The conductor is \textbf{running} after the train.''} \system similarly responds to other clear but plausible discrepancies, such as confusing one character for another, misstating a spatial relation, or describing an object with a property that conflicts with the scene.

Alex does not always take the cue up. If his next utterance still has the conductor walking, the correction path flags the same error again and the salience level that \system{} keeps for actions rises by one step: the Motion Lines return with longer streaks, and on the following cycle, once the streaks are at their longest, with wider bursts of movement, so that the cue becomes harder to overlook while its meaning stays exactly the same. If the description has not changed by the time every parameter of the animation is at its maximum, \system{} changes angle rather than volume and shows the same action through a different animation, flipping the conductor around, which starts from its own defaults. Once Alex says that the conductor is running, the level falls by one step, and it keeps falling each time he describes an action correctly: should an action error be flagged again later, the cue he sees is fainter than the first one was. The levels are updated at every utterance and take effect at the next cycle, so that a gap flagged for the first time is cued at its default and escalates only if it persists.

\begin{figure*}[t]
    \centering
    \includegraphics[width=\textwidth]{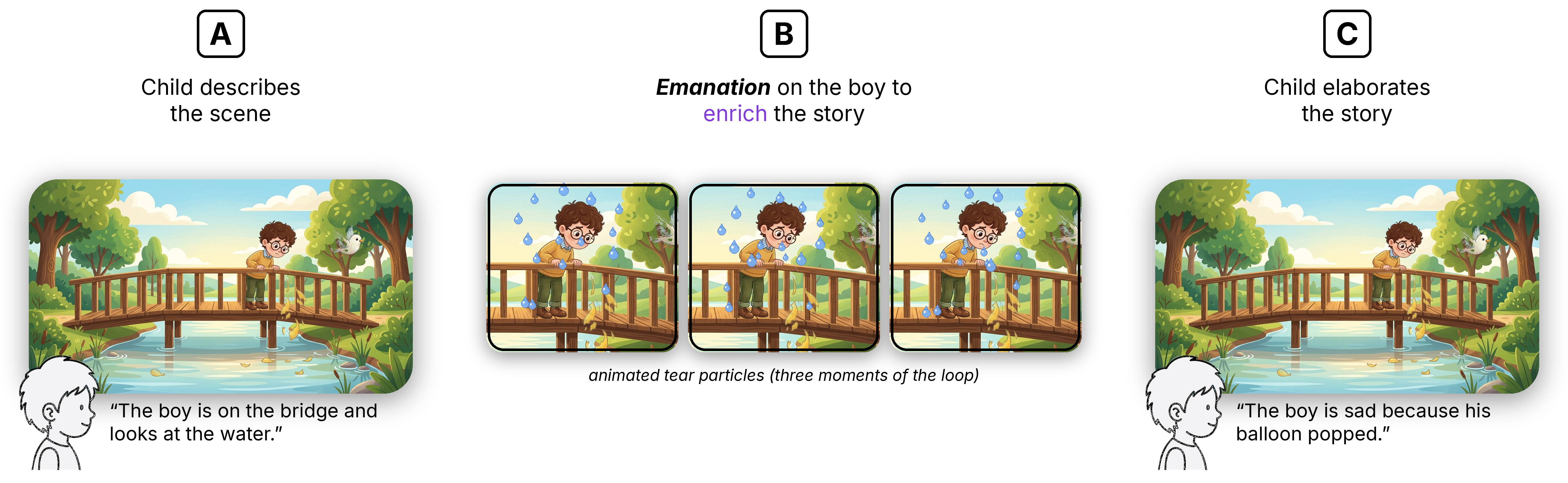}
    \caption{Enriching the story. (A) Alex describes the scene correctly but says nothing about how the boy feels. (B) \system{} plays Emanation on the boy: tear particles make his sadness perceptible without supplying a word. (C) Alex elaborates in his own words.}
    \Description{Three panels labeled A, B, and C. Panel A, child describes the scene: a scene from The Balloon Seller, a boy with glasses leaning on the railing of a wooden bridge over a pond and looking down at the water, with trees around him and a white bird on the right; a line drawing of a child says, The boy is on the bridge and looks at the water. Panel B, Emanation on the boy to enrich the story: three square close-ups of the boy at three moments of the animation loop, with blue tear-shaped particles floating around his head in a different arrangement in each frame. Panel C, child elaborates the story: the same scene as panel A; the child says, The boy is sad because his balloon popped.}
    \label{fig:ux-enrichment}
\end{figure*}

\subsection{Enriching the Story}
More often, Alex's narration is consistent with the scene but leaves some narrative opportunities unexpressed. In \textit{The Balloon Seller}, he reaches the scene where the boy leans over a bridge after his balloon has popped (Figure~\ref{fig:ux-enrichment}). Alex says, \textit{``The boy is on the bridge and looks at the water.''} Nothing in this description is incorrect, but it does not express how the boy feels: an opportunity of the kind \emph{what are they like?} \system responds with an \textbf{Emanation} animation: small tear particles appear around the boy, drawing attention to his sadness without supplying a verbal prompt or the word \textit{``sad.''} Alex can interpret the cue and elaborate in his own words, for example, \textit{``The boy is sad because his balloon popped.''}

Enrichment cues surface narrative possibilities rather than prescribe a continuation: \system considers what Alex has already expressed and points at elements he has not, such as character properties, internal states, plans, or causal relations, and he remains free to decide whether and how to take the cue up. As he begins producing particular kinds of content on his own, \system{} shifts toward other opportunities and lets its cues for what he now does alone fade.

%% file: 05_system.tex
\section{System Architecture}
\label{sec:architecture}

\begin{figure*}[t]
\centering
\includegraphics[width=\textwidth]{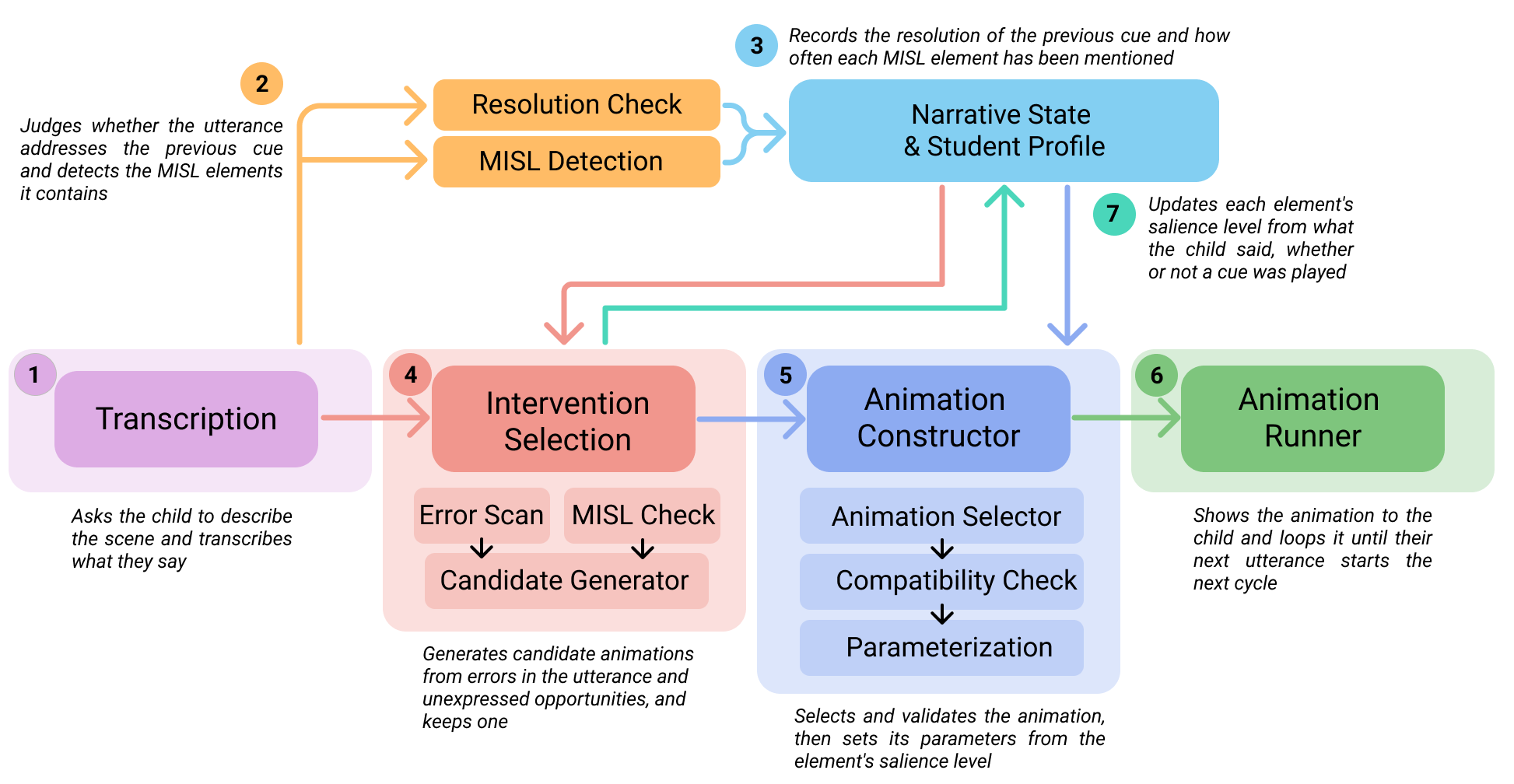}
\caption{Overview of the live interaction loop. Each cycle begins when the child releases the push-to-talk control and the utterance is transcribed (1). Two concurrent model calls judge whether the utterance addresses the previous cue and detect the MISL elements it contains (2), and the narrative state and student profile record both (3). The intervention selection stage scans the utterance for errors and the scene for unexpressed MISL elements, reading mention counts and resolution history from the profile, and keeps at most one candidate, corrections first (4). The animation constructor selects and validates an animation for that candidate and sets its parameters from the salience level the profile holds for its element (5). The animation runner plays the animation until the child speaks again (6). At the end of the cycle, the selection stage writes the updated salience levels back to the profile, whether or not a cue was played (7).}
\Description{Block diagram of the live interaction loop with seven numbered steps. Along the bottom, four boxes connected by arrows from left to right: Transcription (1), Intervention Selection (4), Animation Constructor (5), and Animation Runner (6). Intervention Selection contains Error Scan and MISL Check feeding a Candidate Generator; Animation Constructor contains Animation Selector, Compatibility Check, and Parameterization in sequence. Above, an arrow from Transcription leads to two boxes, Resolution Check and MISL Detection (2), which feed a box labeled Narrative State and Student Profile (3). The profile is connected to Intervention Selection by arrows in both directions and to Animation Constructor by a downward arrow; the return arrow from Intervention Selection to the profile is step 7. Italic notes under each stage summarize its role: transcribing what the child says, judging whether the utterance addresses the previous cue and detecting its MISL elements, recording the resolution and how often each element has been mentioned, generating candidate animations from errors and unexpressed opportunities and keeping one, selecting and validating the animation and setting its parameters from the element's salience level, showing the animation until the next utterance starts the next cycle, and updating each element's salience level whether or not a cue was played.}
\label{fig:pipeline}
\end{figure*}

In this section we describe how the system operationalizes the design space during storytelling. \system operates in two phases. An offline \textit{scene-preparation} phase produces the visual assets and narrative metadata needed to reason about and manipulate each illustrated scene. A \textit{live interaction} phase then processes each child utterance, updates the child’s interaction history, generates correction and enrichment candidates, selects and parameterizes an appropriate animation, and executes it on the relevant scene elements.

\subsection{Scene Preparation}
\label{sec:scene-preparation}

\begin{figure*}[t]
\centering
\includegraphics[width=\textwidth]{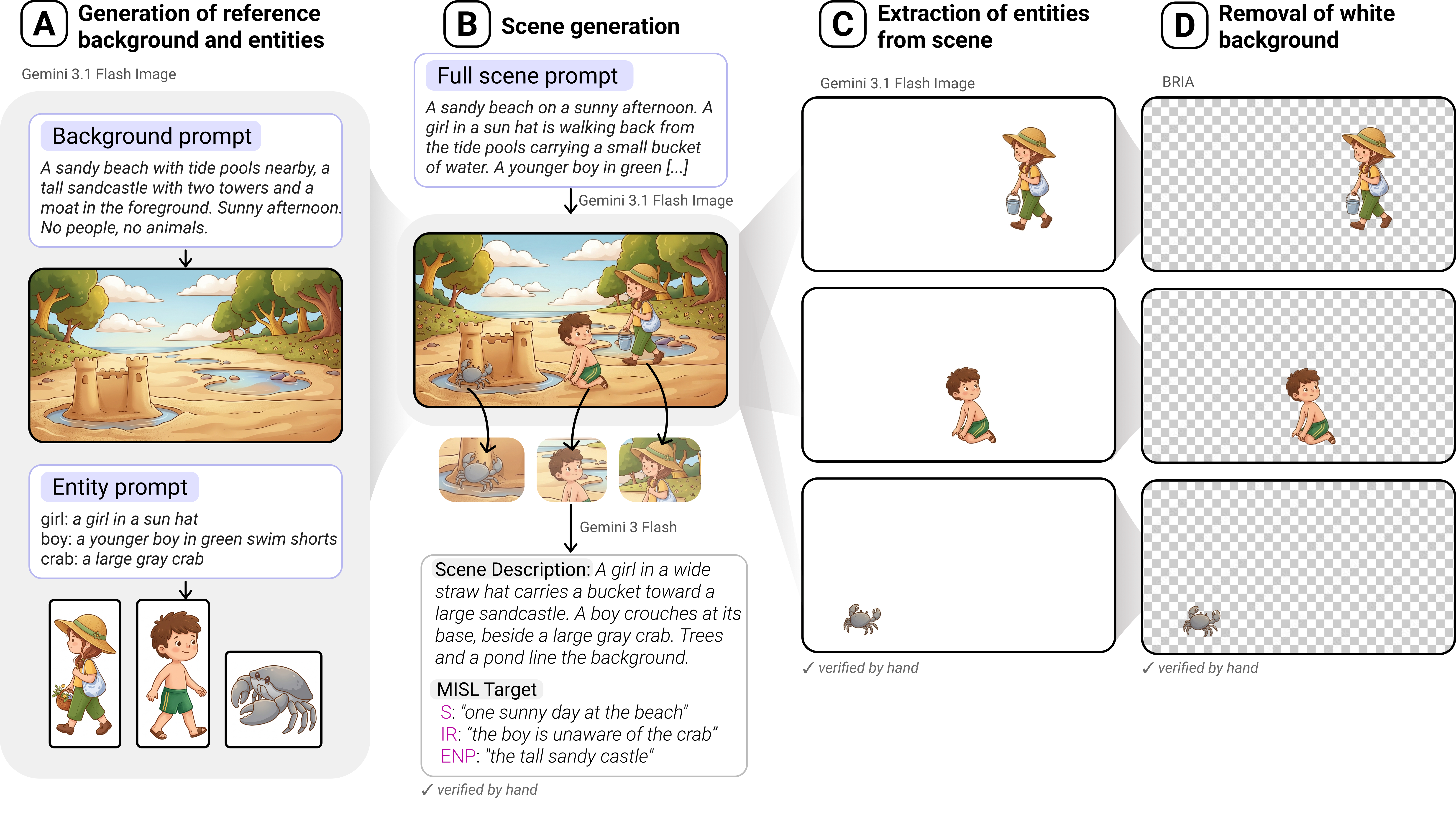}
\caption{Scene preparation pipeline. (A)~From hand-written prompts, Gemini 3.1 Flash Image renders each recurring entity on a white background and a background-only illustration for each scene. (B)~The same model composes the full scene from the background, the entity references, and the full-scene prompt; Gemini 3 Flash then analyzes the composed scene to produce its reference description and the MISL elements it affords. (C)~Gemini 3.1 Flash Image, used as an image editor, isolates each entity in place by whiting out everything else. (D)~BRIA removes the white background, yielding transparent layers aligned with the original illustration. Every output is checked and corrected by hand.}
\Description{Four panels, left to right. Panel A, Generation of reference background and entities, produced by Gemini 3.1 Flash Image: a Background prompt card ("A sandy beach with tide pools nearby, a tall sandcastle with two towers and a moat in the foreground. Sunny afternoon. No people, no animals.") points to an illustration of an empty beach with a sandcastle and tide pools; an Entity prompt card listing three entities (girl: a girl in a sun hat; boy: a younger boy in green swim shorts; crab: a large gray crab) points to three separate drawings of the girl, the boy, and the crab on white backgrounds. Panel B, Scene generation: a Full scene prompt card ("A sandy beach on a sunny afternoon. A girl in a sun hat is walking back from the tide pools carrying a small bucket of water. A younger boy in green [...]") points, via an arrow labeled Gemini 3.1 Flash Image, to the composed scene, which shows the girl carrying a bucket toward the sandcastle while the boy kneels beside it and the crab digs at its base; the background and the three entity drawings from Panel A also feed into this scene. Three arrows from the scene point to cropped views of the crab, the boy, and the girl. An arrow labeled Gemini 3 Flash leads from these crops to an annotation box containing a Scene Description (a short text describing the girl, the sandcastle, the boy, the crab, and the trees and tide pools in the background) and three MISL Target examples (S: "one sunny day at the beach"; IR: "the boy is unaware of the crab"; ENP: "the tall sandy castle"), marked verified by hand. Panel C, Extraction of entities from scene, produced by Gemini 3.1 Flash Image: three images of the same size as the scene, each showing a single entity (girl, boy, crab) at its original position on an otherwise white image, marked verified by hand. Panel D, Removal of white background, produced by BRIA: the same three images with the white replaced by a transparent checkerboard, marked verified by hand.}
\label{fig:preparation}
\end{figure*}

Before a storytelling session begins, each scene goes through an offline preparation pipeline that produces both the visual representation required for animation and the narrative metadata required for language analysis. A central requirement is to preserve the coherence of a complete storybook illustration while making individual scene elements independently manipulable at runtime. We therefore represent each scene as a background together with transparent layers for its characters and objects. This representation allows the animations of Section~\ref{sec:designspace} to target a single entity, a pair or group of entities, or the scene as a whole without regenerating the illustration during interaction.

Figure~\ref{fig:preparation} summarizes how we produce this representation. Each story starts from hand-written prompts: one per recurring entity, and for each scene one for the background alone and one for the full composition. An image-generation model (Gemini 3.1 Flash Image~\cite{gemini31flashimage}) first renders each entity individually on a white background and each scene's background without any entity, then composes the full scene from the background, the entity references, and the full-scene prompt, so that characters keep a consistent identity across scenes and inherit the background's lighting and palette. The same model, used as an image editor, then isolates each entity from the composed scene by replacing everything else with white while keeping the entity at its original position and the image at its original dimensions, and a background-removal model (BRIA~\cite{bria_rmbg2}) turns the result into a transparent layer; each entity receives a stable identifier (e.g., \textit{boy}, \textit{balloon}) that forms the scene's entity registry. Finally, a vision-language model (Gemini 3 Flash~\cite{gemini3flash}) annotates the composed scene with a reference description of its visible entities, actions, properties, locations, and relations, which the correction path later uses to detect discrepancies, and with the MISL elements the scene affords, illustrated by concrete examples (a park scene might afford Setting through ``sunny day'' and Internal Response through ``the boy looks lonely''), which constrain enrichment to what the illustration can support. Every generated output is verified and corrected by hand. We prepared 20 scenes for the children's study (four stories of five scenes each; Section~\ref{sec:eval-children}) and 100 additional scenes for the pipeline-validation and interpretability studies (Sections~\ref{sec:eval-pipeline} and~\ref{sec:eval-interpret}).

\subsection{Live Interaction Loop}
\label{sec:live-loop}

Each interaction cycle begins when the child releases the push-to-talk control. Audio is streamed to the server over a WebSocket connection and transcribed with Google Cloud Speech-to-Text (Chirp 3). The utterance then goes through the steps shown in Figure~\ref{fig:pipeline}, which instantiate the design space: the system identifies a gap and its form, chooses an animation for that kind of gap, parameterizes it, and plays it. A full cycle completes in a median of 5.2 seconds (mean 6.5\,s; 95th percentile 15.5\,s), most of it spent in the model calls of steps 2 and 4; selection, validation, and dispatch together take less than a second.

\paragraph{(2,3) Updating the student profile.}
\label{sec:profile}

\system maintains a profile of the child across the session: how often they have spontaneously produced each of the 15 MISL elements, the names they give to characters (e.g., \texttt{boy} $\rightarrow$ ``Pete''), and whether they addressed each previous animation. Two concurrent Gemini 3 Flash calls update it after every utterance. The first judges whether the utterance makes a reasonable attempt to address the previous animation, if any (\textit{resolved}, \textit{unresolved}, or \textit{not applicable}); the criterion is deliberately generous and does not require a particular wording. The second detects which MISL elements the utterance contains. The profile also holds two salience levels per MISL element, one for corrections and one for enrichment, which persist across sessions, drive the parameterization of step 5, and are written back at step 7.

\paragraph{(4) Generating candidates.}
\label{sec:candidates}
Two paths then run concurrently, each producing at most one candidate: a form (discrepancy or opportunity), a kind of gap, target entities, an animation, and a rationale. The \emph{correction} path compares the utterance with the verified scene description, guided by the correction intents of the 20 animations. It is deliberately conservative: it flags only clear conflicts with visible content and a restricted set of unambiguous grammatical errors, ignoring plausible interpretations, creative elaborations, and deviations typical of the age range. The \emph{enrichment} path first selects, by deterministic rules, a MISL element that the scene affords and the child has rarely produced (macrostructure before microstructure, with preference for elements the child previously left unaddressed), then asks the model which animation and target would make that element salient. These rules are implementation heuristics that limit repetitive feedback, not clinical cutoffs. If neither path finds a gap, the scene stays still.

\paragraph{(5) Selecting, validating, and parameterizing.}
\label{sec:parameterization}
When both paths produce a candidate, correction takes priority: a clear discrepancy is addressed before a new opportunity is introduced. The chosen candidate is validated against the scene before anything reaches the child: the animation must accept the number of targets (\texttt{entity}, \texttt{duo}, \texttt{group}, or \texttt{scene}), otherwise another animation for the same kind of gap is tried, and target identifiers must exist in the entity registry, otherwise they are discarded. The animation's parameters (e.g., dim strength for Spotlight, particle count for Emanation) are then set from the salience level of the element being addressed. Each animation declares an ordered list of its numeric parameters, and each parameter's default sits at the middle of its range (Figure~\ref{fig:spec}). At level zero, every parameter takes its default. Each step above zero moves the first parameter of the list one increment toward its maximum, and once it is there the following steps move the next parameter, so that the intent stays the same while the cue becomes harder to miss along one dimension at a time. Steps below zero walk the same list toward the minimums: a cue for an element the child has started producing on their own recedes gradually rather than snapping back to its default and ends up fainter than it was at first. When every parameter of the current animation is at its maximum and the level continues to increase, the selector changes the angle rather than volume and switches to another animation that addresses the same element (Section~\ref{sec:designspace}).

\paragraph{(6) Playing the animation.}
\label{sec:execution}
The client receives the animation, its targets, parameters, and duration over WebSocket and runs it on the layered scene. Transformations apply to entity layers, using bounds precomputed from the extracted masks; overlays such as particles, bubbles, and marks are generated procedurally; and the runner restores the scene at every frame, so the illustration itself is never altered. The animation loops until the child presses to talk again, which starts the next cycle.

\paragraph{(7) Writing back the salience levels.} Once the animation has been dispatched, the selection stage writes this cycle's update to the profile, whether or not an animation was played: the profile records what the child does, not what the system showed, and the control condition of Section~\ref{sec:eval-children} runs the same state machine. The correction level of an element rises by one step when the correction path flags an error on it, falls by one step when the child produces the element without error, and stays unchanged when the utterance does not mention it, since a child who moves on to another part of the scene has neither corrected nor persisted. The enrichment level rises by half a step when the element was the selected opportunity of the previous cycle and this utterance does not produce it, and falls by a full step whenever the child produces it, cued or not: an omission may be a legitimate narrative choice and is pressed more gently than an error that contradicts the scene. The update takes effect at the next cycle, so that a gap flagged for the first time is cued at its default and escalates only if it persists.


%% file: 06_evaluation.tex
\section{Evaluation}

We evaluate \system{} in three stages that move from system to human. We first assess whether the pipeline selects appropriate animations for a given discrepancy (Section~\ref{sec:eval-pipeline}). We then assess whether observers can interpret those animations without prior instruction (Section~\ref{sec:eval-interpret}). We finally observe how children use \system{} in a real storytelling session (Section~\ref{sec:eval-children}). All three studies were reviewed and approved by the institutional review board of the first author's institution; adult participants gave informed consent online, and for the children's study, parents gave written consent and children gave verbal assent. The first two evaluations rely on online studies with adult participants on Prolific. A first study combined both tasks: the interpretability task (Block~1) always preceded the pipeline validation task (Block~2) to prevent Block~2 from priming participants with explicit descriptions of the pipeline's reasoning. That study also identified three animations whose intended meaning did not come through; we redesigned them and ran a second, shorter interpretability study on the redesigned versions, which we report at the end of Section~\ref{sec:eval-interpret}. 

\subsection{Pipeline Validation}\label{sec:eval-pipeline}

This evaluation targets the selection stage of the pipeline: given a scene and an utterance, does the system identify the right discrepancy and choose an appropriate scaffolding act? It does not assess the visual form of the animations themselves, which is the object of Section~\ref{sec:eval-interpret}. To construct stimuli, we generated 200 synthetic child utterances using Claude Opus 4.6~\cite{claudeopus46}, each paired with a scene description and a narrative context. The utterances were designed to span all eight categories of the design space and to include both factual errors and enrichment opportunities. The first two authors reviewed every utterance and its scene context to confirm that each one presented an unambiguous ground-truth discrepancy. We then ran each utterance through the live pipeline and recorded the animation it selected together with the rationale it produced.

All 200 pipeline decisions were used as stimuli in Block~2 of the first Prolific study. Each stimulus showed participants the scene, the child's utterance, and the rationale the system produced for its decision. Participants rated on a 5-point Likert scale whether the system's response was appropriate for the situation. Each participant saw 10 stimuli (stratified across categories and balancing corrections and suggestions); each stimulus was rated by 4 participants (3 for one stimulus). 80 participants completed Block~2, yielding 799 ratings.

\begin{table*}[t]
\centering
\caption{Pipeline validation ratings per decision type (80 participants, 799 ratings on 200 decisions, 5-point scale): mean with a 95\% confidence interval obtained by bootstrap over stimuli, share of ratings at 4 or above, and mean per condition.}
\label{tab:pipeline-validation}
\small
\renewcommand{\arraystretch}{1.15}
\begin{tabular}{@{}lllrlrrr@{}}
\toprule
\textbf{Category} & \textbf{ID} & \textbf{Gap flagged by the decision} & $n$ & \textbf{Mean [95\% CI]} & $\geq$\textbf{4 (\%)} & \textbf{Corr.} & \textbf{Sugg.} \\
\midrule
Identity & I1 & wrong or unmentioned entity or setting & 32 & 4.19 [3.62, 4.59] & 78 & 3.94 & 4.44 \\
 & I2 & wrong or missing name or role & 32 & 3.78 [3.34, 4.28] & 62 & 3.88 & 3.69 \\
\addlinespace[2pt]
Count & C1 & miscounted or uncounted group & 32 & 4.47 [4.09, 4.81] & 91 & 4.44 & 4.50 \\
 & C2 & entity mentioned that is not in the scene & 32 & 2.84 [2.31, 3.47] & 41 & 2.62 & 3.06 \\
 & C3 & entity absent from this scene but in the story & 32 & 2.12 [1.56, 2.66] & 25 & 1.44 & 2.81 \\
\addlinespace[2pt]
Property & P1 & wrong or missing visible attribute & 32 & 4.22 [3.97, 4.47] & 75 & 4.38 & 4.06 \\
 & P2a & hot object or shame missed & 32 & 3.81 [2.87, 4.56] & 72 & 3.50 & 4.12 \\
 & P2b & cold object or coldness missed & 32 & 3.78 [3.34, 4.22] & 62 & 3.75 & 3.81 \\
 & P2c & joy missed or misattributed & 32 & 3.44 [2.88, 3.88] & 56 & 3.50 & 3.38 \\
 & P2d & affection missed or misattributed & 32 & 4.06 [3.69, 4.44] & 84 & 3.81 & 4.31 \\
 & P2e & anger missed or misattributed & 32 & 4.22 [3.94, 4.50] & 81 & 4.38 & 4.06 \\
 & P2f & fear missed or misattributed & 32 & 3.88 [3.03, 4.50] & 69 & 3.56 & 4.19 \\
\addlinespace[2pt]
Action & A1 & wrong or unmentioned movement & 32 & 4.03 [3.56, 4.47] & 69 & 4.12 & 3.94 \\
 & A2 & wrong or unmentioned non-movement action & 32 & 3.88 [3.25, 4.44] & 72 & 3.56 & 4.19 \\
\addlinespace[2pt]
Space & S1 & hidden element denied or unmentioned & 32 & 3.75 [3.31, 4.12] & 66 & 3.44 & 4.06 \\
 & S2 & entity misplaced or location unmentioned & 32 & 4.41 [4.03, 4.75] & 88 & 4.19 & 4.62 \\
\addlinespace[2pt]
Time & T1 & past event not marked as past & 32 & 3.78 [3.44, 4.12] & 62 & 3.69 & 3.88 \\
 & T2 & ongoing or upcoming process described as instant & 32 & 3.19 [2.84, 3.56] & 47 & 3.38 & 3.00 \\
\addlinespace[2pt]
Relation & R1 & entities together described as apart & 32 & 3.78 [3.19, 4.41] & 69 & 3.25 & 4.31 \\
 & R2 & entities apart described as together & 32 & 4.06 [3.41, 4.56] & 72 & 3.75 & 4.38 \\
 & R3 & cause wrong, reversed, or unmentioned & 31 & 3.55 [2.81, 4.22] & 61 & 3.53 & 3.56 \\
\addlinespace[2pt]
Discourse & D1 & speech misattributed or unmentioned & 32 & 3.19 [2.50, 3.81] & 47 & 2.50 & 3.88 \\
 & D2 & feeling or intention contradicted or unmentioned & 32 & 4.09 [3.78, 4.34] & 75 & 3.88 & 4.31 \\
 & D3 & key event wrong or unmentioned & 32 & 4.44 [4.06, 4.66] & 94 & 4.62 & 4.25 \\
 & D4 & grammatical error or word to elaborate & 32 & 3.56 [2.91, 4.12] & 62 & 3.00 & 4.12 \\
\midrule
\textbf{All} & & & 799 & 3.78 & 67 & 3.60 & 3.96 \\
\bottomrule
\end{tabular}
\end{table*}

\begin{figure*}[t]
\centering
\includegraphics[width=0.7\textwidth]{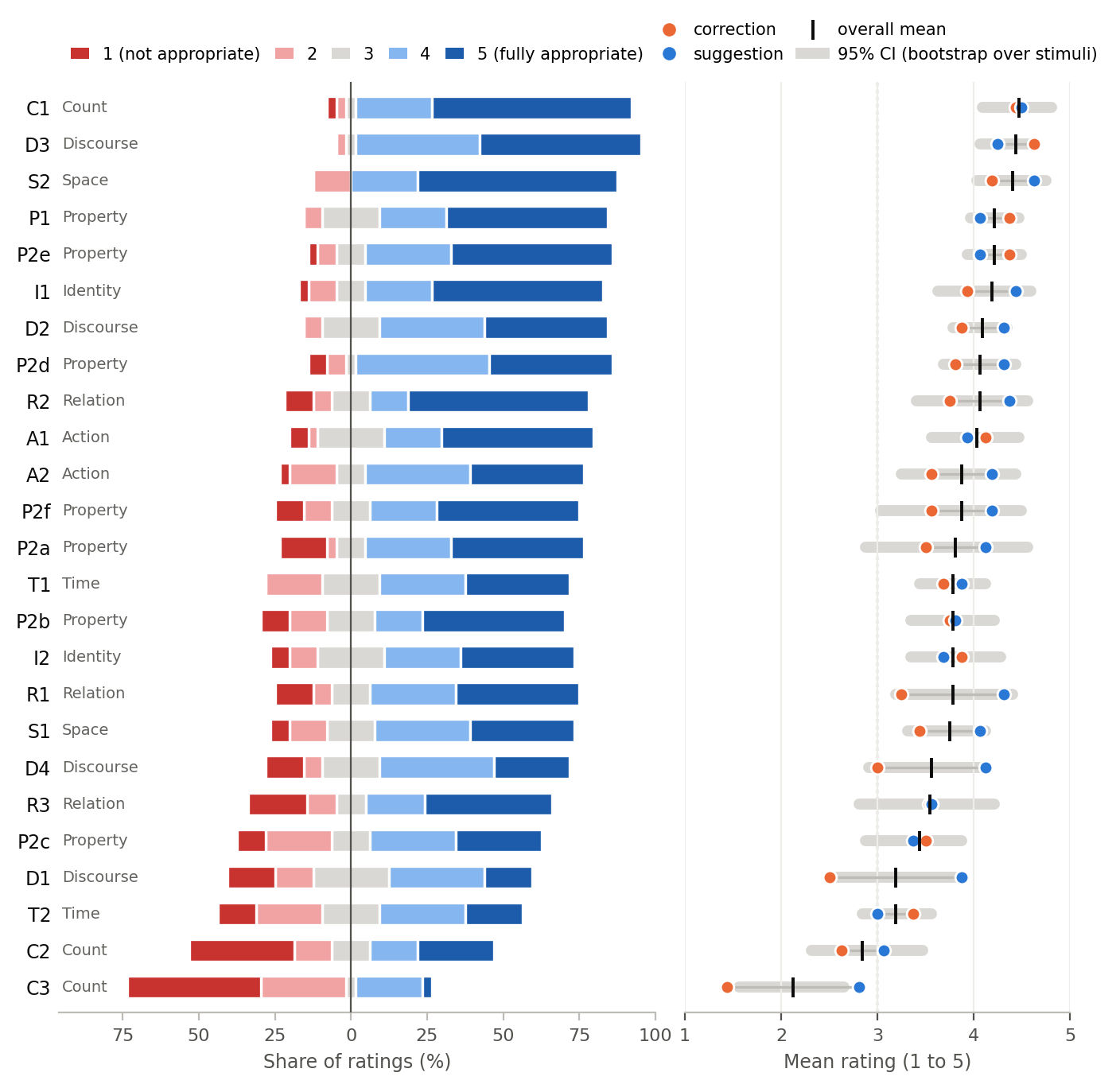}
\caption{Pipeline validation ratings per decision type, sorted by mean (80 participants, 799 ratings). Left: distribution of the ratings on the 5-point scale, centered on the neutral point. Right: mean rating for corrections and suggestions, overall mean, and 95\% confidence interval obtained by bootstrap over stimuli.}
\Description{Two-panel chart with one row per decision type, 25 rows sorted from the highest mean rating at the top (C1, Count) to the lowest at the bottom (C3, Count), each labeled with its identifier and category. Left panel, share of ratings in percent: a diverging stacked bar per row, centered on the neutral rating; ratings of 1 and 2 extend to the left in dark and light red, ratings of 3 sit in gray at the center, and ratings of 4 and 5 extend to the right in light and dark blue. Bars at the top are almost entirely blue; the red part grows toward the bottom, and the last two rows, C2 and C3, are mostly red. Right panel, mean rating from 1 to 5: for each row, an orange dot for corrections, a blue dot for suggestions, a black tick for the overall mean, and a gray horizontal band for the 95 percent confidence interval obtained by bootstrap over stimuli. The blue dot sits to the right of the orange dot for most rows; the gaps are largest for D1, R1, and D4, where corrections are rated well below suggestions, and C3 has the lowest mean, with its correction dot near 1.5.}
\label{fig:pipeline-validation}
\end{figure*}

\textit{Results.} The overall mean rating was 3.78 out of 5 (median: 4), and two thirds of all ratings fell at 4 or above, indicating that participants generally found the pipeline's decisions appropriate. Table~\ref{tab:pipeline-validation} reports the ratings per decision type and Figure~\ref{fig:pipeline-validation} their distribution. At the category level, decisions about Space (mean 4.08), Identity (3.98), Action (3.95), and Property (3.92) were rated highest, followed by Discourse (3.82) and Relation (3.80); Time (3.48) and Count (3.15) received the lowest ratings. The picture is sharper at the level of the individual decision types, that is, the detection rationale attached to each animation (Section~\ref{sec:animations}). The best-rated decisions flag a miscounted or uncounted group (C1, 4.47), a key event described wrongly or left unmentioned (D3, 4.44), a misplaced entity (S2, 4.41), and a missed or misattributed anger (P2e, 4.22). The lowest-rated flag a reference to an entity that belongs to the story but is absent from the current scene (C3, 2.12), an entity mentioned by the child that is not in the scene at all (C2, 2.84), an ongoing or upcoming process described as instantaneous (T2, 3.19), and speech attributed to the wrong character (D1, 3.19). The two lowest cases share a property that explains them: the discrepancy concerns something that is not in the picture, so the rationale asks the participant to trust the system's knowledge of the story rather than what they can see. Decisions grounded in visible content, which make up the large majority of the pipeline's decisions, were rated favorably across all categories. Suggestions (mean 3.96) were rated higher than corrections (mean 3.60), suggesting that the pipeline's enrichment decisions were perceived as more natural than its error-correction decisions.

\textit{Decisions rated as inappropriate.} We examined the 32 decisions (16\%) whose mean rating fell below 3. In 7 of them (3.5\% of the 200 decisions) the pipeline had misrecognized the gap. Five corrections addressed a discrepancy other than the one the utterance contained: one rationale flagged as absent a maple tree that is in the scene, another read a tree bearing the leaves of several seasons as a scene set in the future. Two suggestions proposed an enrichment that the scene does not support: a cause invented for a broken fence, and a teddy bear singled out as out of place in a toy box where the misplaced object is a shoe. One further stimulus was a construction error on our part: the displayed utterance no longer contained the error that the rationale addressed. In the remaining 24 decisions, spread across 14 decision types, the rationale named the discrepancy the utterance contained and the low rating bears on the decision itself; they include four of the six low-rated decisions about an entity absent from the scene (C3) and two of the four grammatical corrections (D4).

\textit{False positive analysis.} The ratings above evaluate the pipeline's decisions when it intervenes, not whether it flags valid utterances. We therefore randomly sampled two utterances per animation from the 200 stimuli (50 in total), corrected each one by hand so that it accurately described the scene, and ran them through the pipeline. Of these, 44 passed without intervention and 6 produced a spurious correction: two were hallucinated errors, and four were misreadings of the narrative context, where an utterance consistent with the scene was read as contradicting its description. We did not observe any false negatives during the review (errors present in the utterance that the pipeline failed to detect), so the pipeline errs, when it does, on the side of intervening.

\subsection{Animation Interpretability}\label{sec:eval-interpret}

The pipeline validation confirms that the system selects the right scaffolding act, but the animation must also communicate its intended meaning to the viewer. If adults cannot interpret an animation without instruction, children will not be able to either. We assessed this in two studies: one covering the whole design space, and a second on the three animations that the first singled out for redesign.

\textit{Study 1: the full design space.} We created 100 scenes: 25 animations (the 20 animations of the design space, with Emanation tested separately for each of its 6 particle variants) paired with 4 scene variants each. Each scene was used in two conditions, correction and suggestion, yielding 200 stimuli. Each stimulus presented a picture of the scene, a narration prompt describing what a child has just said, then a video of the system's animated response, and asked participants what they would change or add to their narration. In the correction condition, the five options were the target interpretation, two partial-credit options (the right element with the wrong act, or the right act on the wrong element), and two distracters; in the suggestion condition, the target and four distracters. The first two authors reviewed all stimuli to confirm that the target entity and intended narrative gap were unambiguous. We recruited 80 participants through Prolific (age $\geq$ 18, first language English, normal color vision, desktop only), not informed of the study's connection to children's storytelling. The 200 stimuli were organized into 8 counterbalanced lists of 25, one stimulus per animation, each assigned to 10 participants.

\textit{Results.} Across all trials, overall accuracy was 49.6\%, well above the 20\% chance level. At the category level, Action (58.0\%) and Property (55.8\%) were the best-interpreted categories; Identity (31.9\%) and Time (41.4\%) were the lowest, both still above the 20\% chance level. Corrections (53.3\%) were interpreted more accurately than suggestions (46.0\%), consistent with corrections highlighting a visible contradiction while suggestions draw attention to an absence. We tested whether each animation exceeded chance using a one-sided binomial test with Bonferroni correction ($\alpha = .002$); 22 of the 25 animations significantly exceeded chance. Accuracy was highest for Emanation-Tears (P2f, 73.8\%). Three animations did not reach significance, each for an identifiable reason: Nametag (I2, 10.1\%), the only animation below chance, because participants interpreted the floating label as flagging a mistake rather than inviting naming; Ghost Outline (C3, 25.9\%), whose translucent silhouette of an absent entity was read as a present entity being highlighted; and Reveal (S1, 33.3\%), whose fade of the occluding entity did not convey that something was hidden behind it. 

\textit{Redesign.} We redesigned these three animations, keeping their position in the design space. Nametag became \emph{Silhouette}: the entity turns into a flat black silhouette in place, holds, then snaps back, asking who or what it is. Ghost Outline became \emph{Missing Piece}: a jigsaw-shaped hole opens where the absent entity would be, exposing a flat backing, while the matching piece, bearing the entity's silhouette, hovers at the frame edge. Reveal became \emph{Peek}: the occluding entity lifts like a flap hinged on one edge, exposing what is behind it, then closes back. All three follow the requirements of the design space: whole-layer transformations or procedural overlays, no added text or symbol, and a return to the initial state at the end of each loop.

\textit{Study 2: the redesigned animations.} We re-ran the interpretability task on the three redesigned animations with the same scenes, narration prompts, and options as Study~1, so that only the video differed. Each participant saw the 12 target stimuli (3 animations $\times$ 4 scenes, one condition per scene, counterbalanced across 4 lists) plus 3 filler stimuli taken unchanged from Study~1 (one animation each at a high, medium, and low Study~1 accuracy) to check that the two samples were comparable. We recruited 40 new participants with the same criteria, excluding all Study~1 participants, yielding 469 target and 119 filler responses. The comparison with Study~1 is therefore between subjects.

\textit{Results.} Accuracy improved for all three animations. Silhouette reached 21.8\% (vs.\ 10.1\% for Nametag, Fisher's exact test $p = .031$), Missing Piece 32.7\% (vs.\ 25.9\%, $p = .30$), and Peek 39.0\% (vs.\ 33.3\%, $p = .48$). Pooled over the three animations, accuracy rose from 23.2\% to 31.1\% ($p = .028$), and Missing Piece and Peek now significantly exceed chance ($p < .001$, one-sided binomial). Silhouette more than doubled its accuracy and, although it does not yet exceed chance, its errors are now almost exclusively partial-credit answers that identify the right entity (81\% of correction trials) rather than distracters: participants look where the animation points, and what remains open is which act to perform on that entity. Peek's remaining difficulty is concentrated in a single scene: when the occluder is a small entity (a cat on a chair), 7.9\% of participants identified the hidden element, against 42 to 63\% on the three other scenes, which suggests a parameter constraint (a minimum occluder size) rather than a flaw in the animation itself. These gains are, if anything, conservative: the Study~2 sample was less accurate than the Study~1 sample on the identical filler stimuli (47.1\% vs.\ 63.0\%), while response times and video replays were the same across the two studies.


\begin{figure*}[t]
\centering
\includegraphics[width=\textwidth]{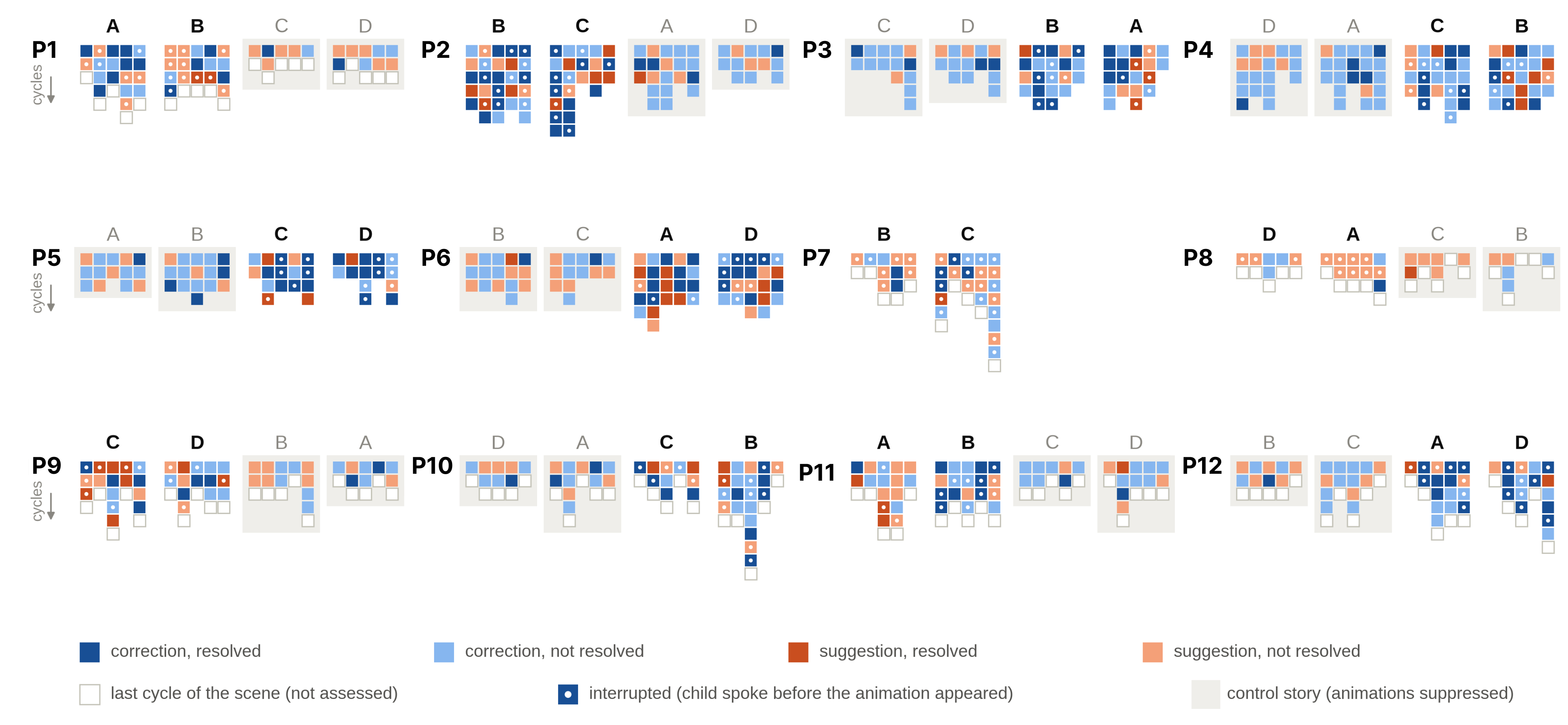}
\caption{Every pipeline cycle of the twelve sessions. Each block is one story told by one child, with the four stories in the order the child told them: the five scenes of the story are the columns of the block and the cycles of each scene run from top to bottom, one square per utterance. Hue gives the type of decision the pipeline made on that utterance (blue for a correction, orange for a suggestion); a dark fill marks a gap the child addressed on the next utterance, a light fill one that went unaddressed, and an empty square the last cycle of a scene, which no utterance follows. A white dot marks an animation the child interrupted by speaking again before it appeared. A gray plate marks a control story, where the pipeline ran identically but displayed nothing. P7 told only the two animated stories.}
\Description{A grid of small colored squares arranged in eight groups labeled P1 to P8, four groups per row over two rows. Each group holds four blocks labeled with a story letter, A to D; each block is five columns of squares of unequal height, between one and nine squares tall. Squares are dark blue, light blue, dark orange, light orange, or white with a gray outline, and some carry a white dot. Two blocks in every group, and all four blocks in none, sit on a light gray plate; the plated blocks contain almost only light squares, while the unplated blocks mix dark and light squares. The group P7 has two blocks only, both unplated, with white dots on most of their squares. A legend below the grid names the five square styles and the gray plate.}
\label{fig:sessions}
\end{figure*}

\subsection{Storytelling with Children}\label{sec:eval-children}

We conducted a within-subjects study with 12 children (three aged 7, four aged 8, two aged 9, one aged 10, and two aged 11; mean = 8.6, SD = 1.4), recruited via a flyer circulated by email. All used English at school; 7 had been formally diagnosed with a speech or language difficulty, and 5 were identified by their parents as presenting language difficulties. We deliberately used the broad term ``language difficulties'' in our recruitment materials rather than a clinical label in order to include children whose difficulties have not been formally assessed. Each child told 4 stories (2 with animations, 2 in a control condition where the pipeline ran identically but suppressed all visual output), assigned via a Latin square; one child (P7) stopped after the two animated stories, so the analysis covers 24 animated and 22 control stories. Sessions were conducted over Zoom, lasted 38 minutes on average (SD = 10.3), and the interaction was logged from the first onboarding utterance to the last utterance of the last story. Each session began with an onboarding phase, in which the child practiced with two short practice stories to familiarize themselves with the push-to-talk interface and the animations. The option to advance to the next scene, which the tool ties to the coverage of the scene (Section~\ref{sec:ux}), was made available one minute after each scene appeared, so that no child could be held back on a scene. For both conditions, the pipeline logged every discrepancy it identified and whether the child's next utterance in the same scene addressed it (a resolution); in the control condition, resolutions reflect spontaneous self-corrections. Figure~\ref{fig:sessions} shows every cycle of every session as one square, colored by the type of decision and by whether the child resolved it; the contrast between animated and control stories is visible before any count.

\textit{Results.} In the animation condition, the pipeline ran 20.6 cycles per story, and every cycle flagged a gap: 13.2 corrections and 7.4 suggestions per story. Of the animations it triggered, 11.9 per story were displayed and 8.7 were interrupted, the child having pressed the push-to-talk button again before the animation appeared. In the control condition, the pipeline ran 14.7 cycles per story (9.6 corrections and 5.1 suggestions) with all visual output suppressed. Children thus produced more utterances with animations, and animated stories lasted slightly longer (7.9 vs.\ 7.1 minutes). The number of distinct MISL elements per story was similar in both conditions (12.9/15 with animations, 12.2/15 in control). The difference appeared in resolutions. With animations, children addressed 197 of the 425 gaps that were followed by another utterance in the same scene (46\%), against 34 of 265 (13\%) in control; per story, 8.2 gaps were resolved out of 17.7 assessed, against 1.5 out of 12.0. Ten of the eleven children who completed both conditions resolved a larger share of gaps with animations than without (Wilcoxon signed-rank test on per-child rates, $W = 1$, $p = .002$); the remaining child (P8) resolved one gap in each condition. Corrections were resolved more often than suggestions in both conditions (53\% vs.\ 34\% with animations, 17\% vs.\ 5\% in control), consistent with the interpretability results. Across the animated stories, resolution was highest for Discourse (59\%), Count (56\%), and Property (56\%) decisions and lowest for Identity (33\%) and Relation (30\%). At the level of individual animations, Interjection (D4, 71\%) and Alert (D3, 67\%) were resolved most often and Silhouette (I2, 20\%) least often, mirroring its low interpretability score; Spotlight (I1), the most frequently played animation (22\% of all animations), was resolved in 39\% of cases. 

\begin{figure*}[t]
\centering
\includegraphics[width=\textwidth]{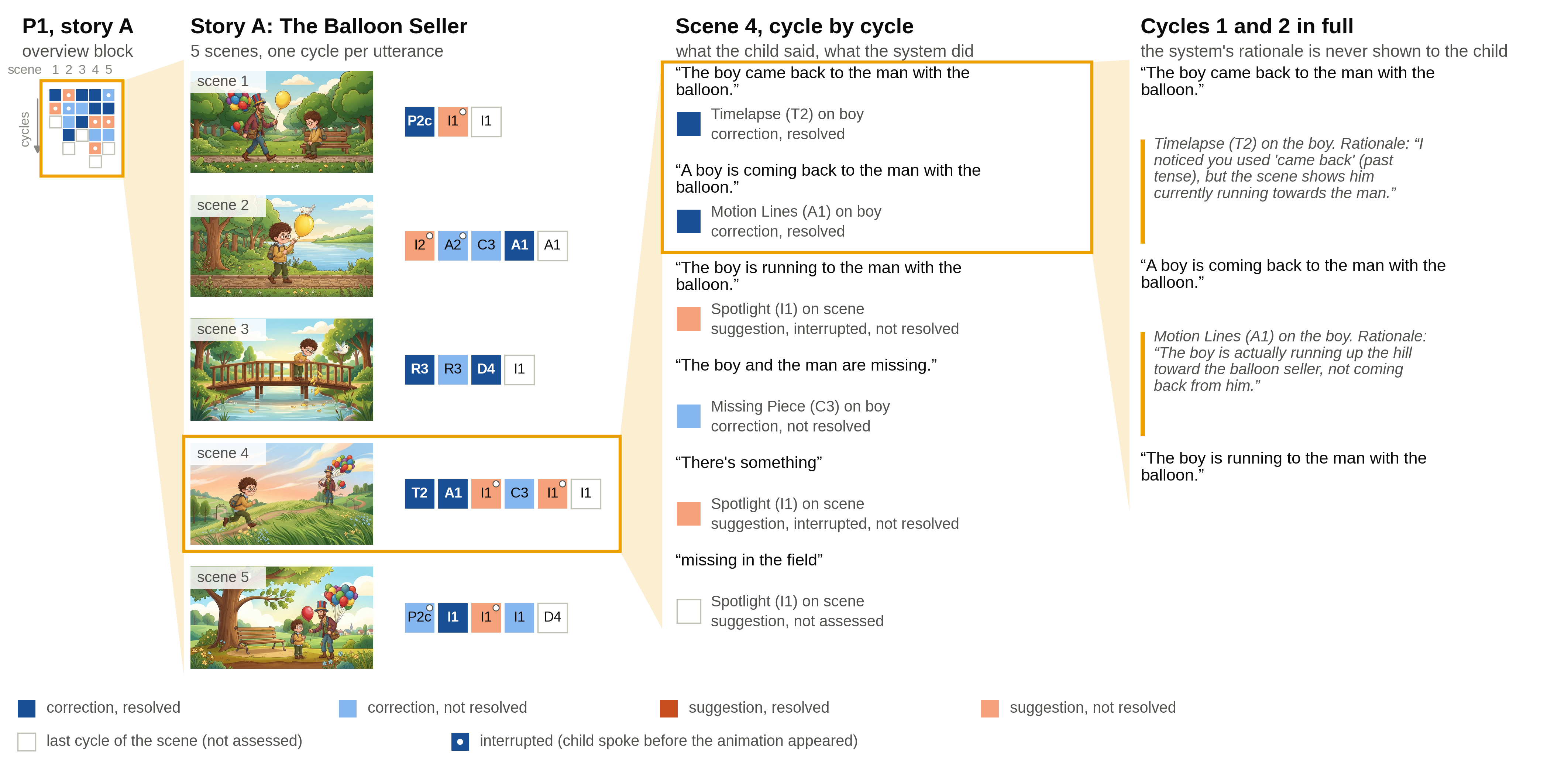}
\caption{One episode at four levels of detail, from the overview to the utterances. Left: the P1, story A block of Figure~\ref{fig:sessions}. Second column: the five scenes of the story, each with the animation the pipeline played at each cycle, labeled by its identifier; dark squares are decisions the child addressed on the next utterance, and a small dot marks an interrupted animation. Third column: scene 4 cycle by cycle, with the child's utterance, the animation, its target, and the outcome. Right: the first two cycles in full, with the rationale the pipeline produced for each decision. The rationale is internal to the pipeline and is never shown to the child. Highlighted bands link each level to the part of the next level it expands.}
\Description{Four columns of increasing detail, connected by pale orange bands. The first column is a small block of colored squares with an orange frame. The second column lists five scenes of a story, each with a small illustration of a boy and a balloon seller and a row of labeled squares such as T2, A1, I1, C3. The row for scene 4 is framed in orange. The third column lists six cycles of scene 4, each with a quoted utterance of the child, a colored square, the name of the animation with its identifier, its target, and whether the correction or suggestion was resolved; the first two cycles are framed in orange. The fourth column repeats those two cycles with the child's three utterances in quotation marks and, between them, the pipeline's rationale for each animation in italics. A legend at the bottom names the square styles.}
\label{fig:varifocal}
\end{figure*}

Figure~\ref{fig:varifocal} follows one episode from the overview block down to the utterances: the story, its five scenes, the cycles of one scene, and the exchange in full. In that episode, a child (P1, age 7) said ``The boy came back to the man with the balloon.'' The system played Timelapse (T2) to flag the tense error; the child corrected to ``A boy is coming back to the man with the balloon.'' The system then played Motion Lines (A1) to highlight that the boy is running, not simply coming back; the child responded ``The boy is running to the man with the balloon.'' Over two cycles and two animation types, the description moved from an inaccurate past-tense summary to a correct present-tense action. In a second episode, a child (P8) said ``He's getting closer to his home'' while the scene showed the boy in the park receiving a new balloon from the seller. The system played Alert (D3) on the boy; the child reformulated to ``He is really excited that he has a balloon now,'' replacing an invented event with the boy's actual state. The logs also show where the interaction breaks down. Animations were interrupted in 42\% of cases overall (the dotted squares in Figure~\ref{fig:sessions}), and in 29 of the 39 cases for P7, whose transcripts interleave the child's answers with an adult's questions: the turns arrived faster than the pipeline could respond, and most animations were never seen.

Three takeaways follow. First, the animations did not change how much of the MISL rubric the children covered, but they changed what the children did with the gaps the pipeline identified: the resolution rate more than tripled for corrections and suggestions alike. Second, corrections were resolved more readily than suggestions in both conditions, consistent with the interpretability results, but the animations increased the resolution rate of suggestions from 5\% to 34\%, so that an animation helps a child act on an absence and not only on an error. Third, the pace of the exchange matters: an animation that arrives after the child has started the next utterance is lost, which puts a premium on the latency of the pipeline and on the timing of the display relative to the child's turns.

%% file: 07_discussion.tex
\section{Discussion}\label{sec:discussion}

\subsection{\system as Attentional Scaffold}

Our central question was whether a system can draw a child's attention to a narrative gap without biasing what they say. The children's study says it can: with the same pipeline running in both conditions, the share of identified gaps that children addressed more than tripled when the animations were shown, and they addressed them in their own words, since no animation ever supplied one. Three findings say more about how this works.

First, an animation that says \emph{where to look} does more than deliver a message. In the logs, a single Spotlight on a misidentified character led a child to correct the character, describe the setting, and notice an entity that no animation had targeted. A question such as ``Who is that?'' would have obtained the correction but framed the answer; the animation redirected the gaze and left the rest of the scene open, the attentional counterpart of the finding that pedagogical instruction narrows exploration~\cite{bonawitz2011double}. Interpretability and attentional affordance are therefore two distinct properties of an animation. Silhouette shows the difference: it does not yet exceed chance in the forced-choice task, but 81\% of its errors identify the right entity; participants look where it points before they know what it asks. Evaluations of attentional feedback should measure both properties, and a design may legitimately favor the second.

Second, pointing at what is present is easier than pointing at what is absent. Corrections were interpreted more accurately than suggestions (53.3\% vs.\ 46.0\%), children resolved them more readily, and the three animations that needed redesign all referred to something not in the picture: an unnamed identity, an absent entity, a hidden one. The pipeline validation shows the same asymmetry from the other side, since the two lowest-rated decision types concerned entities absent from the scene. Yet the gaps that matter most for language development are omissions rather than errors, and raters found the pipeline's enrichment decisions more natural than its corrections (3.96 vs.\ 3.60). The design challenge of attentional scaffolding is therefore concentrated in one place: making an absence perceptible.

Third, when a cue is not taken up, changing the angle appears more productive than turning up the volume. The pipeline escalates by intensity (Section~\ref{sec:parameterization}), but the most productive sequences in the logs were those in which successive cycles addressed the same gap through different animations, a Timelapse on a tense error followed by Motion Lines on the underlying action, for instance. Practitioners of language intervention do the same when they recast rather than repeat a prompt that did not land. The many-to-many mapping between MISL elements and kinds of gaps (Section~\ref{sec:gap-forms}) makes such reformulation possible by construction; our implementation does not yet exploit it deliberately, and the observation is a hypothesis to test.

\subsection{Design Guidelines for Guided Storytelling Animations}

The two adult studies and the children's study suggest six guidelines for systems that use animation to direct attention during a language task.

\textit{\textbf{DG1:} Transform what is there before adding what is not.} The best-interpreted kinds of gaps, Action and Property, are served by animations that transform visible content, and all three redesigns moved in this direction: a label became a transformation of the entity (Silhouette), a translucent overlay became a hole in the scene (Missing Piece), a fade became a physical lift (Peek). Overlays work when they borrow a convention children already read, as tears and speech bubbles do, and fail when they invent one, as the empty label of Nametag did.

\textit{\textbf{DG2:} Make absence visible through presence.} To point at something missing, alter something present, a hole where the entity should be or a lifted occluder, rather than drawing what is not there. The same holds for the system's reasoning: raters trusted decisions grounded in visible content and distrusted those that relied on the story's off-screen knowledge.

\textit{\textbf{DG3:} Set perceptual floors on parameters.} Peek was read correctly on three scenes and almost never on the one whose occluder was small. Each parameter needs bounds tied to the geometry of the scene, such as a minimum occluder size or a minimum motion amplitude relative to entity size, so that the animation stays apprehensible~\cite{tversky2002}.

\textit{\textbf{DG4:} Keep the intent stable, vary the intensity, then change the angle.} Escalation should first raise salience without changing meaning and, if the gap persists, switch to another animation that addresses the same narrative element through a different kind of gap. It should also run in reverse: once the child produces the element, the cue should recede step by step rather than reset, so that the scaffold fades instead of stopping.

\textit{\textbf{DG5:} Correct conservatively, enrich generously.} A spurious correction contradicts a child who was right, and the false-positive analysis shows where the risk lies, in hallucinated errors and misread context. Enrichment carries no such cost and was judged natural; thresholds should reflect this asymmetry.

\textit{\textbf{DG6:} Leave the words to the child.} No animation in the design space displays a word the child could repeat; the one that shows text, Interjection, shows the child's own word. This constraint is what separates attentional scaffolding from a hint.

\subsection{Broader Utility across Clinic, Classroom, and At Home}

Children tell stories from pictures in several settings, and each calls for a different workflow. In the clinic, \system would add a second, silent channel to verbal scaffolding rather than replace it: a clinician could enable only the kinds of gaps a child is working on (Action and Property for a child consolidating verbs and descriptive language, for instance), cap the intensity of the cues, or review the session afterwards. Because the pipeline detects MISL elements in every utterance and records whether each cue was addressed, it produces as a by-product the progress-monitoring data that MISL was designed to capture~\cite{gillam2017, israelsen2022} and that automated scoring has begun to approximate~\cite{jones2019, fox2022}. In the classroom, where teachers rarely have training in narrative assessment, a summary of which kinds of gaps recur for which child would let a teacher target small-group work without scoring transcripts. At home, the priority is independence: the salience levels already reduce support as the child improves, in line with intervention practice~\cite{spencer2020narrative}, and the pace of that fading becomes the parent's control.

In every setting, what \system can scaffold is bounded by what its scenes afford: a panel in which no character shows an emotion cannot support an Internal Response, and a story without a problem cannot support an Initiating Event. Generative models rendered our scenes and extracted their layers, but the stories, the arc across panels, and what each panel was meant to afford were authored by us and verified by hand, and we expect this to remain true: the narrative potential of a scene is a pedagogical decision, not a rendering task. The next step is an authoring workflow in which an SLP or teacher writes the story and marks the elements each panel should afford, the pipeline generates scenes to that specification, and the verification steps become an editing interface, open as well to illustrations already used in practice once their entities are extracted as layers.

The mechanism itself is not specific to storytelling or to children with language difficulties: wherever a speaker describes a visual scene and a system can compare the two, in second-language picture description or show-and-tell for instance, an animation can point at what was missed or misstated, and the three-layer structure of the design space (what kind of gap, in which form, made perceptible how) is a template for such vocabularies.

\subsection{Limitations}

The interpretability studies used adults as proxies for children. The choice is conservative: participants received no onboarding, were not told the animations were meant for children, and decoded each one cold, whereas children saw them repeatedly, in the flow of a story, after two practice stories. But a seven-year-old's exposure to the conventions we build on differs from an adult's, and a forced-choice task measures how an animation is decoded, not how it redirects attention in the moment. The children's study observed the second effect without measuring it separately; gaze tracking during narration would be the natural complement.

The children's study was a single session per child. It shows that children take up the animations and elaborate their narration in response, not whether this carries over to unscaffolded storytelling once the novelty of a moving picture has worn off. Narrative intervention is evaluated over weeks with progress-monitoring probes~\cite{gillam2017, israelsen2022}; the corresponding evaluation of \system is a longitudinal deployment with MISL scores on unscaffolded narratives as the outcome and the pipeline's logs as the process measure. The silent control isolates the contribution of the visual channel; a third arm with verbal prompting would test the complementarity we hypothesize.

On the system side, entities are indivisible layers, so a character cannot smile, turn its head, or change posture, which is why Internal Response is scaffolded through emanata rather than facial expression, as an illustrator would do it; articulated characters from future generation models would add a family of animations that the design space accommodates without change. Scene preparation still involves human verification, which makes the annotations reliable but bounds the number of stories. \system also supports English only: the rubric, transcription, and prompts are language-specific, the animations are not, so another language would be a direct test of how far the design space travels.

%% file: 08_conclusion.tex
 \section{Conclusion}

We presented \system, an approach that scaffolds children's picture-based storytelling by animating the scene elements that their narration has left out or misdescribed, so that a narrative gap becomes visible without a word being supplied. Its design space characterizes narrative gaps as eight unanswered questions, each of which can arise as a discrepancy or as an opportunity, and provides 20 animations, grounded in the conventions of animated media, to make them perceptible. A real-time pipeline detects both forms of gap in each utterance, selects an animation for the kind of gap it finds, and adjusts its intensity to the child's history, in a median of 5.2 seconds. Adult raters judged the pipeline's decisions appropriate, 22 of 25 animations were interpreted above chance without instruction, and a redesign of the remaining three improved all of them. In a within-subjects study with 12 children, the animations more than tripled the share of identified gaps that children addressed, in their own words. These results suggest that directing attention is a scaffold in its own right, and that animation offers AI-supported learning tools a way to hint at what matters without saying what to say.
